\documentclass[final]{siamltex}

\usepackage{epsfig}
\title{Statics of point Josephson junctions in a micro strip line.}

\author{J.G. Caputo\thanks{Laboratoire de Math\'ematiques,
                 Institut de Sciences Appliquees, 
                 B.P. 8, 76131 Mont-Saint-Aignan
                 Cedex, France and Laboratoire de Physique 
                 theorique et modelisation,
                 Universit\'e de Cergy-Pontoise and C.N.R.S.}
        \and L. Loukitch \thanks{Laboratoire de Math\'ematiques,
                 Institut de Sciences Appliquees,
                 B.P. 8, 76131 Mont-Saint-Aignan
                 Cedex, France.}}

\begin{document}

\maketitle

\begin{abstract}
We model the static behavior of point Josephson junctions in a micro strip 
line using a 1D linear differential equation with delta 
distributed sine non-linearities. 
We analyze the maximum current $\gamma_{max}$ crossing the 
micro strip for a given magnetic field $H$. In particular we establish
its periodicity and analyze how it is affected by
the geometry, length, type of current feed, position and area of the junctions.
For small currents, which is the rule in practice, we show that
$\gamma_{max}$ can be obtained by a simple formula, the magnetic approximation.
This model is in excellent agreement with measurements obtained for
real devices.
\end{abstract}



\pagestyle{myheadings}
\thispagestyle{plain}
\markboth{J.G. Caputo and L. Loukitch}{Static of Josephson junctions in a micro strip line.}

\section{Introduction}

The coupling of two low $T_c$ superconductors across a thin oxide layer is 
described by the Josephson equations \cite{josephson}.
\begin{equation}\label{e1.1}
V=\Phi_0\frac{d\phi}{dt},~~~I=sJ_c\sin(\phi)~,
\end{equation}
where $V$ and $I$ are, respectively, the voltage and current across 
the barrier, $s$ is the contact surface, $J_c$ is the critical current 
density and $\Phi_0=\hbar/2e$ is the reduced flux quantum. 
These two Josephson relations together with Maxwell's equations imply 
the modulation of DC current by an external magnetic field in the static 
regime and the conversion of AC current into microwave radiation 
\cite{Barone,Likharev}. Other applications include Rapid single flux quantum
logic electronics\cite{Likharev} and microwave signal mixers used in
integrated receivers for radio-astronomy\cite{Salez}.
In all these systems there is a characteristic 
length which reduces to the Josephson length, $\lambda_J$, the ratio 
of the electromagnetic flux to the quantum flux $\Phi_0$ for standard junctions.

For many applications and in order to protect the junction, Josephson junctions are 
embedded in a so called microstrip line which is the capacitor made
by the overlap of the two superconducting layers. This is the so-called 
"window geometry" where the phase difference between the top and bottom layer 
satisfies an inhomogeneous 2D damped driven sine Gordon equation \cite{cfv95} 
resulting from Maxwell's equations and the Josephson constitutive relations
(\ref{e1.1}). The damping is due to the normal electrons and the driving 
through the boundary conditions with an external current or 
magnetic field applied to the device.

Even in the static regime the 2D problem is complicated because of the 
multiplicity of solutions due to the sine term. However 
flux penetration occurs along the 
direction of the magnetic field so one direction dominates the other.
A quantity measured by experimentalists is the maximum (static) current 
$I_{\rm max}(H)$ that can cross the device for a given magnetic 
field $H$. This gives informations on the quality of the junctions.
An important issue is how defects in the coupling will affect this
maximum current.  In particular high $T_c$ superconductors can be described as 
Josephson junctions where the
critical current density is a rapidly varying function of the position, 
due to grain boundaries. Fehrenbacher et al\cite{fgb92} calculated 
$I_{\rm max}(H)$ for such disordered long Josephson junctions and 
for a periodic array of defects. Experiments were also done by 
Itzler and Tinkham on large 2D disordered junctions \cite{it9596}.
However the overall picture is complex and it is difficult from the
curve $I_{\rm max}(H)$ to obtain geometric information on the junction.
The analysis of such a 2D problem \cite{cftv03} provided bounds on the
gradient of the solution that were independent of the area of the junctions
so that little information could be obtained on $I_{\rm max}(H)$. However the
study \cite{cftv03} proved the existence of solutions and the convergence
of the Picard iteration to obtain them.

Small junctions of length $w_i < \lambda_J$ are easier to study and 
lead to the well known $I_{\rm max}(H)=\sin(H w_i) / H$\cite{Barone}. Two
such junctions are commonly associated to form a Superconducting Quantum
Interference Device (SQUID) now routinely used to measure magnetic fields. 
More junctions can be used to form arrays \cite{uclocr95} that can bear more critical
current and are more flexible than a long junction because the area of the
junction components and their position can be varied. When the junctions 
are closer than $\lambda_J$, such arrays behave as a long junction and
could be used as microwave generators. Almost all models are discrete
lumped models where the effect of the space between junctions is neglected. 
In particular the interaction of the junctions through this passive region
has always been neglected. This makes it difficult to describe junctions
of different areas, placed non uniformly in the microstrip.
This is why up to now mostly equidistant and identical junctions have been considered.
\begin{figure}
\centerline{\epsfig{file=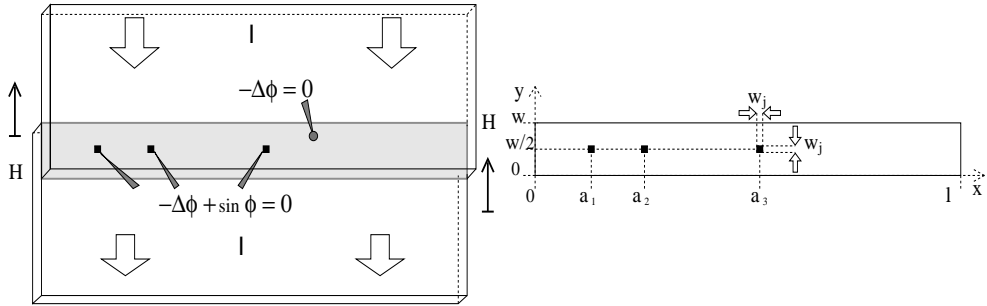,height=4 cm,width=13 cm,angle=0}}
\caption{The left panel shows the top view of a superconducting 
microstrip line containing three Josephson junctions,
$H,I$ and $\phi$ are respectively the applied magnetic field, current and the
phase difference between the two superconducting layers. The phase difference
$\phi$ between the two superconducting layers satisfies 
$-\Delta \phi=0$ in the linear part and 
$-\Delta \phi+\sin(\phi)=0$ in the Josephson junctions. The right panel
shows the associated 2D domain of size $l \times w$ containing $n=3$ junctions
placed at the positions $y=w/2$ and $x=a_i,~i=1,n$.}
\label{f1} \end{figure} 
To overcome these difficulties we recently introduced a continuous/discrete 
model that preserves the 
continuity of the phase and its normal gradient across the junction interface
and where the phase is assumed constant in the junctions. The 1D dynamics \cite{cl} of one junction
in a cavity revealed that the junction could stop waves across the cavity
or enhance them throughout. In \cite{cg04} this model was used
to calculate $I_{\rm max}(H)$ for a miss-aligned SQUID in a 2D cavity.
Nonuniform arrays of junctions that are generalized SQUIDs have 
been produced and analyzed in particular by Salez et al at the
Observatory of Paris\cite{Salez} and our analysis is in excellent agreement
with the measured $I_{\rm max}(H)$.

In this article we will concentrate on the 1D static problem and show that it allows
for an in depth analysis that was out of reach in the general 2D case. In particular
we will show the properties of $I_{\rm max}(H)$, its periodicity, 
its regularity, the relation between different types of current 
feeds and how it is affected
by the position of the array in the microstrip. In addition we introduce 
and justify the so-called magnetic approximation where many details
of $I_{\rm max}(H)$ can be controlled. 
Specifically in section two we introduce our model and 
give preliminary analytical
results in section three. Section four details the 
intrinsic properties of the maximal current as a function of 
the magnetic field: its periodicity,
the relation between the inline and overlap current feed and the simple 
magnetic approximation.
Section five introduces two numerical ways to solve the problem. In the 
sixth section, we study a SQUID and examine the effect of a little
difference between the junction parameters and we compare this to the experiment. 
Section seven deals with devices with more junctions, there we analyze the effect of 
separating one junction from the others and show the agreement with 
the experimental results.

\section{The model}

The device we model shown in Fig.~\ref{f1} is a so-called microstrip
cavity (grey area in Fig.~\ref{f1}) between two superconducting layers.
Inside this microstrip there are regions where the oxide layer is
very thin ($\sim$ 10 Angstrom) enabling Josephson coupling between the top and
bottom superconductors. The dimensions of the microstrip are about
100 $\mu$m in length and 20 $\mu$m in width. The phase difference between 
the top and bottom superconducting layers obeys in the static regime 
the following semilinear elliptic partial differential equation \cite{cfv95}
\begin{equation}\label{2dsg}
-\Delta\varphi+g(x,y)\sin\varphi = 0,
\end{equation}
where $g(x,y)$ is 1 in the Josephson junctions and 0 outside. This formulation 
guarantees the continuity of the normal gradient of $\varphi$, the 
electrical current on the junction interface. The unit of space is the 
Josephson length $\lambda_J$, the ratio of the flux formed with the
critical current density and the surface inductance to the flux quantum
$\Phi_0$. 

The boundary conditions representing an external current input $I$
or an applied magnetic field $H$ (along the y axis) are
\begin{eqnarray}
\left. \frac{\partial \varphi}{\partial y}\right|_{y=0}=-\frac{I }{2l}\nu~, &
\left. \frac{\partial \varphi}{\partial y}\right|_{y=w}=\frac{I }{2l}\nu~,
\nonumber\\[-1.5ex]
\label{bc} \\[-1.5ex]
\left. \frac{\partial \varphi}{\partial x}\right|_{x=0}=H-\frac{I}{2w}(1-\nu)~,&
\left. \frac{\partial \varphi}{\partial x}\right|_{x=l}=H+\frac{I}{2w}(1-\nu)~, 
\nonumber
\end{eqnarray} 
where $0\leq \nu \leq 1$ gives the type of current feed. The case
$\nu = 1$ shown in Fig.~\ref{f1} where the current is only applied
to the long boundaries $y=0,w$ is called overlap feed while
$\nu=0$ corresponds to the inline feed.

We consider long and narrow strips containing a few small junctions
of size $w_j \times w_j$ placed on the line $y=w/2$ and centered on $x=a_i, ~~i=1,n$
as shown in Fig.~\ref{f1}. We then search $\varphi$ in the form
\begin{equation}\label{e2.2.1}
\varphi(x,y)=\frac{\nu I }{2L}\left(y-\frac{\omega}{2}\right)^2 + 
\sum_{n=0}^{+\infty}\phi_n(x) \cos\left(\frac{n\pi y}{w}\right),
\end{equation}
where the first term takes care of the $y$ boundary condition.
For narrow strips $w<\pi$, only the first transverse 
mode needs to be taken into
account \cite{cfgv96,bcf02} because the curvature of $\varphi$ due
to current remains small. Inserting (\ref{e2.2.1}) into (\ref{2dsg})
and projecting on the zero mode we obtain the following 
equation for $\phi_0$
where the $0$'s have been dropped for simplicity
\begin{equation}\label{e2.3}
-\phi^{\prime \prime} + g\left(x,{w\over 2}\right){w_j\over w} \sin \phi = 
\nu \frac{\gamma}{l},
\end{equation}
where $\gamma=I/w$ and the boundary conditions
$\phi^{\prime}(0)=H-(1-\nu)\gamma/2$, 
and $\phi^{\prime}(l)=H+(1-\nu)\gamma/2$. 

As the area of the junction is reduced, the total
Josephson current is reduced and tends to zero. 
To describe small junctions where the phase variation
can be neglected but that can carry a significant current,
we introduce the following function $g_h$
\begin{equation}\label{gh}
g_h(x) = {w_j\over 2 h} ~~~{\rm for}~ a_i-h<x<a_i+h,~~g_h(x) =0~~
~~{\rm elsewhere},\end{equation}
where $i=1,\dots n$. 
In the limit $h\rightarrow 0$ we obtain our final delta function
model \cite{cl}
\begin{equation}\label{e3.1}
-\phi^{\prime\prime}+\sum_{i=1}^n d_i \delta(x-a_i)\sin\phi = \nu j,
\end{equation}
where 
\begin{equation}\label{defdj}
d_i = {w_j^2 \over  w},~~  j={\gamma \over l} 
\end{equation}
and the boundary conditions are 
\begin{equation} \label{bcon} 
\phi^{\prime}(0) = H-(1-\nu)\gamma/2,   ~~ 
\phi^{\prime}(l) = H+(1-\nu)\gamma/2. 
\end{equation}
This is our continuous/discrete 1D model of a parallel array of 
many point Josephson junctions embedded in micro strip cavity.
It preserves the spatial degrees of freedom in the linear cavity
and the matching conditions at the junction interfaces.

\section{General properties}

The delta function seems to be a theoretical way to approach the problem.
Nevertheless we will show that it provides an excellent agreement with
experiments, in addition to allow simple calculations.
We have the following properties.
\begin{remunerate}
\item Integrating twice (\ref{e3.1}) shows that the solution
$\phi$ is continuous at the junctions $x=a_i, ~i=1,\dots n$.
\item Let $\phi$ be a solution of the equation (\ref{e3.1}), 
then $\phi+2k\pi$ is also a solution. 
\item Almost everywhere, 
$-\phi^{\prime\prime}(x)= \nu \gamma/l$,
so that outside the junctions, $\phi$ is a second degree 
polynomial by parts,
\begin{equation}\label{rem1}
\phi(x) = -\frac{\nu j}{2}x^2 + B_i x + C_i~,~~ \forall x \in ]a_i,a_{i+1}[.
\end{equation}
\item At each junction ($x=a_i$), $\phi^{\prime}$ is not defined, but 
choosing $\epsilon_1 > 0$, and $\epsilon_2 > 0$, we get
$$\lim_{\epsilon_1 \rightarrow 0}\lim_{\epsilon_2 \rightarrow 0} 
\int_{a_i-\epsilon_1}^{a_i+\epsilon_2} \phi^{\prime\prime}(x) dx =
\int_{a_i^-}^{a_i^+}\phi^{\prime\prime}(x) 
dx= \left[\phi^{\prime}(x)\right]_{a_i^-}^{a_i^+}.$$
Since the phase is continuous at the junction $x=a_i$, we obtain:
\begin{equation}\label{rem2}
\left[\phi^{\prime}(x)\right]_{a_i^-}^{a_i^+} = d_i\sin(\phi_i)~,
\end{equation}
with $\phi_i \equiv \phi(a_i)$.
\item Integrating (\ref{e3.1}) over the whole domain, 
$$\left[\phi^{\prime}\right]_0^l= \int_0^l \phi^{\prime\prime} dx =
\sum_{i=1}^n d_i \sin(\phi_i)-\nu \gamma~,$$ and taking into account the
boundary conditions, we obtain
\begin{equation}\label{rem3}
\gamma = \sum_{i=1}^n d_i\sin(\phi_i)~,
\end{equation}
which indicates the conservation of current. Note that
the total current is equal to the sum of the jumps
of $\phi^{\prime}$.
\end{remunerate}

\subsection{Polynomial by part}

Let $\phi$ be a solution of (\ref{e3.1}) and $\phi_1=\phi(a_1)$.
From remark (\ref{rem1}), $\phi$ is a polynomial by parts. We define 
$P_{i+1}(x)$ the second degree polynomial such that $P_{i+1}(x)=\phi(x) 
~~{\rm for}~~ a_{i} \le x \le a_{i+1}$.
Using the left boundary condition we can specify $\phi$ on $[0,a_1]$: 
\begin{equation}\label{rec1}
P_1(x)=-\frac{\nu j}{2}\left(x^2-a_1^2\right)
+ \left(H-\frac{1-\nu}{2}\gamma\right)(x-a_1) + \phi_1~.
\end{equation}
At the junctions (\ref{rem2}) tells us that $\forall k \in \{1,\dots,n\}$,
\begin{equation}\label{jump_a}
P_{k+1}^{\prime}(a_k) - P_k^{\prime}(a_k)= d_k \sin(P_k(a_k)).
\end{equation}
Considering that $\phi^{\prime \prime}=-\nu j$ on each interval, 
the previous relation and the continuity
of the phase at the junction, we can give a first expression for $P_{k+1}$,
\begin{equation}\label{recn}
P_{k+1}(x)=-\frac{\nu j}{2}(x-a_k)^2  + \left[P^{\prime}_k(a_k)+d_k \sin
P_k(a_k) \right](x-a_k)+P_k(a_k).
\end{equation}
Notice that $P_{k+1}(x)$ depends on $P_k(x)$, $\nu$, $j$ 
and $H$. The parameters $\nu$ and $l$ are fixed by the geometry of the device. 
So by recurrence we see that
$\phi$ is entirely determined by the values of $\phi_1$, 
$\gamma$ and $H$.

From (\ref{jump_a}) we can obtain another expression for $P_{k+1}$ 
\begin{equation}\label{jump}
P_{k+1}(x) - P_k(x) = d_k \sin(P_k(a_k))(x-a_k).
\end{equation}
Summing all these relations yields 
\begin{equation}\label{global_P}
P_{k+1}(x) = P_1(x) + \sum_{i=1}^k d_i \sin(P_i(a_i))(x-a_i).
\end{equation}
Polynomials (\ref{rec1}) and (\ref{recn}) show by construction, that the
constants $H$, $j$ and $\phi_1$, determine completely the solution of  
(\ref{e3.1}) if it exists. In same way, we can show that the three 
other constants, $j$, $\phi^{\prime}(a_1)$ and $\phi_1$ fix $\phi$. From
(\ref{global_P}), we give an expression of $\phi$ 
$$\phi(x) = P_1(x) + \sum_{i=1}^n {\cal H}_{\{x \geq a_i\}} d_i \sin(\phi_i)(x-a_i),
$$
where ${\cal H}_{\{x \geq a_i\}} = \left\{ \begin{array}{lr}
1,& x \geq a_i, \\ 0, & x<a_i. \end{array} \right.$ is the Heaviside 
function.

\subsection{An $n$ th order transcendental system}

Another way to solve (\ref{e3.1}) for $\phi$ 
is to write it as a coupled system of $n$ transcendental equations.
For that, we first eliminate the constant term by introducing $\psi$ such that
$$\phi = \psi -\nu {\gamma \over l} {x^2 \over 2}\equiv \psi -f(x)$$
and obtain 
\begin{equation}\label{sg_psi}
-\psi^{\prime\prime}+\sum_{i=1}^n d_i \delta(x-a_i)
\sin(\psi-f(a_i)) = 0,
\end{equation}
with the boundary conditions
$$ \psi^{\prime}(0) = H-(1-\nu)\gamma/2,   ~~ 
\psi^{\prime}(l) = H+(1+\nu)\gamma/2. $$
To simplify the notation we will write $f_i\equiv f(a_i)$ and 
$\psi_i \equiv \psi(a_i)$.
Integrating (\ref{sg_psi}) over the intervals $[0,a_2^-],~[a_1^+,a_3^-],~..$
we obtain the relations
\begin{eqnarray} \label{int_rels}
-[\psi^{\prime}]_{0}^{a_2^-} + d_1\sin(\psi_1-f_1)=0 , \nonumber \\
-[\psi^{\prime}]_{a_1^+}^{a_3^-} + d_2\sin(\psi_2-f_2)=0 , \nonumber \\
-[\psi^{\prime}]_{a_2^+}^{a_4^-} + d_3\sin(\psi_3-f_3)=0 ,\\
-[\psi^{\prime}]_{a_3^+}^{a_5^-} + d_4\sin(\psi_4-f_4)=0 , \nonumber \\
-[\psi^{\prime}]_{a_4^+}^{l} + d_5\sin(\psi_5-f_5)=0 , \nonumber 
\end{eqnarray}
where we have assumed $n=5$ as an example.
Now we can use the fact that $\psi^{\prime \prime}=0$ in the intervals between the
junctions and the boundary conditions to obtain the final system
\begin{eqnarray}\label{system}
H -(1-\nu){\gamma \over 2}- {\psi_2-\psi_1 \over a_2-a_1} + 
d_1\sin(\psi_1-f_1)=0 , \nonumber \\
-{\psi_3-\psi_2 \over a_3-a_2} +  {\psi_2-\psi_1 \over a_2-a_1} + 
d_2\sin(\psi_2-f_2)=0 , \nonumber \\
-{\psi_4-\psi_3 \over a_4-a_3} +  {\psi_3-\psi_2 \over a_3-a_2} + 
d_3\sin(\psi_3-f_3)=0 , \\
-{\psi_5-\psi_4 \over a_5-a_4} +  {\psi_4-\psi_3 \over a_4-a_3} + 
d_4\sin(\psi_4-f_4)=0 , \nonumber \\
-H -(1+\nu){\gamma \over 2} + {\psi_5-\psi_4 \over a_5-a_4} 
+ d_5\sin(\psi_5-f_5)=0 . \nonumber 
\end{eqnarray}
We will use this formulation as well as the
one in the previous subsection to establish properties of the solutions
and solve the problem numerically using Newton's method.

\section{General properties of $\gamma_{max}(H)$ for an $n$ junction array}

The general problem is
\begin{equation}\label{e5.1}
-\phi^{\prime\prime}(x)+\sum_{i=1}^n d_i \delta(x-a_i)\sin(\phi) = \nu j.
\end{equation}
with the boundary conditions
$$ \phi^{\prime}(0) = H-(1-\nu)\gamma/2 , ~~
\phi^{\prime}(l) = H+(1-\nu)\gamma/2 .$$
Experimentalists measure the maximum current $\gamma$ for a given
magnetic field $H$ and plot this as a curve $\gamma_{max}(H)$. To compare 
with real data it is therefore important to compute and analyze this 
quantity.  In this section, we give some properties of the 
$\gamma_{max}(H)$ curve.
In the appendix some analytical estimates on the influence of the
geometry on the maximal current will be presented.

\subsection{Periodicity}

We introduce $$l_j \equiv a_{j+1}-a_j,$$ the distance between two consecutive junctions. 
Let $l_{min}$ be the smallest distance $l_j$. We define the array as 
harmonic if $l_i$ is a multiple of $l_{min}$ for all $i$. 

\begin{figure}
\centerline{\epsfig{file=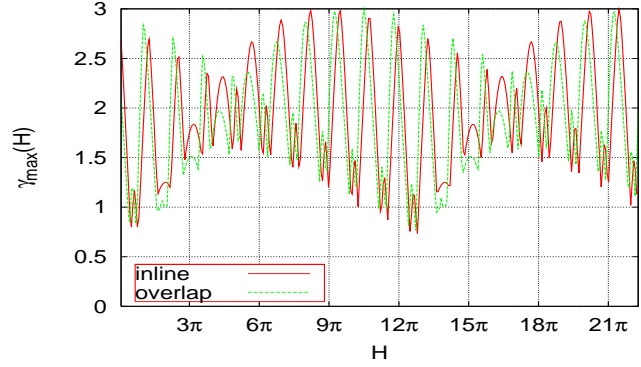,height=9 cm,width=5 cm,angle=270}}
\caption{$\gamma_{max}(H)$ curve for an inline current feed, $\nu=0$ (continuous line)
and overlap feed $\nu=1$ (dotted line) for a three junction unit $\left\{1,5/2,5/2+5/3\right \}$,
with $d_1=d_2=d_3=1$. So $l_1=3/2$ and $l_2=5/3$. }
\label{f2} \end{figure}

\begin{proposition}[{\rm Periodicity of the device}]
\label{pro1} For a harmonic array, the $\gamma_{max}(H)$ 
curve is periodic with a period $2\pi/l_{min}$.
\end{proposition}

\begin{proof}
Let $\phi$ be a solution of (\ref{e5.1}) for a current $\gamma$ and a magnetic 
field $H$. We introduce $f(x)=(2\pi/l_{min})(x-a_1)$ and 
$\psi(x) = \phi(x) + f(x)$. So $\psi$ verifies
\begin{equation}
-\psi^{\prime\prime}(x)+\sum_{i=1}^n \delta(x-a_i)\sin(\psi-f) = \nu j.
\end{equation}
with $\psi^{\prime}(0) = H+2\pi/l_i - (1-\nu)\gamma/2$, and
$\psi^{\prime}(l) = H+2\pi/l_i + (1-\nu)\gamma/2$.  
Since, $f(a_j)=2k\pi$, $\forall i\in \{1,...,n\}$, then $\psi$ is a solution
of (\ref{e5.1}) for $H+H_p \equiv H+2\pi/l_{min}$ and the same $\gamma$, so
$\gamma_{max}(H+H_p) \ge \gamma_{max}(H)$.

Conversely, by subtracting $f$ from a solution associated to $H+H_p $
and a current $\gamma$, we obtain a solution for $H$ and the same 
current $\gamma$ so $\gamma_{max}(H+H_p) \le \gamma_{max}(H)$. From the
two inequalities we get 
\begin{equation}\label{Potd}
\gamma_{max}(H+H_p)=\gamma_{max}(H)~.
\end{equation}
with $H_p = 2\pi/l_{min}$.
\qquad\end{proof}

In the non harmonic case, if the junctions are set such that
$l_j=p_j/q_j$, where $p_j$ and $q_j$ are integers, prime with 
each other, then  $\gamma_{max}$ is periodic with period $H_p$ such that
\begin{equation}\label{periodicity}
H_p = 2\pi \frac{LCM(q_1,...,q_{n-1})}{HCF(p_1,...,p_{n-1})}~,
\end{equation}
see Fig.~\ref{f2}, 
where $LCM$ is the Lowest Common Multiple and $HCF$ the 
Highest Common Factor. 
To prove this write 
$f(x) = p(x-a_1)$ and use again the previous argument.
In Fig.~\ref{f2} we show the $\gamma_{max}(H)$ curve for a three junction unit
such that $l_1=3/2$ and $l_2=5/3$ so that the period 
of $\gamma_{max}(H)$ is $H_p= 2\pi LCM(2,3)/HCF(3,5)=12\pi$. In the following plots
we will only show one period of $\gamma_{max}(H)$.

In the general case, we only have an approximate periodicity of 
$\gamma_{max}(H)$ which can be estimated using (\ref{periodicity}).
Also, real junctions have a finite size which causes 
$\gamma_{max}(H)\rightarrow 0$, when $H\rightarrow +\infty$. Our
model is thus valid as long as the dimensionless magnetic field 
$H$ is not larger than $1/w_j$. 

\subsection{Influence of the position of the junction unit}

In this section, we examine how the position of the set of
junctions in the microstrip (linear domain) will affect the
$\gamma_{max}(H)$ curve.
For an array of junctions placed at the distances $\{a_i, i=1,n\}$, we
define a junction unit as the set $\{l_{i},~i=1,n-1\}$. Then
the array where the junctions are at $\{a_1+c,a_2+c,...,a_n+c\}$ is 
the same junction unit. We define $a_1$ as the position of the junction
unit. The length of the junction unit is $l_b=a_n-a_1$.
The array is centered if $(a_n + a_1)/2= l/2$.

{\it Inline current feed:} ($\nu=0$) 

Then the boundary conditions at the edge of the junction 
unit are 
$\phi^{\prime}(a_1^-) =\phi^{\prime}(0) = H-\gamma/2$, and 
$\phi^{\prime}(a_n^+) = \phi^{\prime}(l) = H+\gamma/2$, independently
of the position of the junction unit.
\begin{proposition}[{\rm Inline junction unit}]
For inline current feed, $\gamma_{max}(H)$ is independent of $a_1$
(the position of the junction unit) and of the length $l$ of the cavity.
\end{proposition}

{\em Proof}. Let $\phi_1(x)$ be a solution of (\ref{e5.1}), for given 
$\gamma$, $H$. Let us change the position of the junction unit to
$a_1+c$ so that the junctions are now placed at $\{a_1+c,a_2+c,...,a_n+c\}$.
It is easy to see that $\phi_2(x)=\phi_1(x-c)$ satisfies the boundary 
conditions and is a solution. This one to one map between $\phi_1$ and 
$\phi_2$ exists for all $c, H$ and $\gamma$ so the two junction units
have the same $\gamma_{\rm max} (H)$.
\endproof

Then the $\gamma_{max}(H)$ curve is independent of the position of junction unit
when $\nu=0$. By the same argument, we can show that $\gamma_{max}(H)$ is
independent of the length $l$ of the circuit (see Fig.~\ref{f4}). This curve 
depends only on the junction unit.

{\it General current feed:} ($0<\nu\leq1$) 

Then the boundary conditions at the edge of the junction unit are: 
$$ \phi^{\prime}(a_1^-) = -\nu j a_1 + H + (1-\nu)\gamma/2,~~ 
\phi^{\prime}(a_n^+) = H - (1-\nu)\gamma/2 + \nu j (l-a_n).$$
Contrary to the inline feed, we cannot shift the phase to find a solution when
the junction unit has been shifted, because now the 
boundary conditions depend on the position of the junction unit. 
Consider the derivative $\phi^{\prime}$ at the 
boundaries of the junction unit. We will compare the curves $\gamma_{max}(H)$ 
for a centered unit and for a non centered unit. For a centered unit, $a_n-a_1=l/2$
so that
$$\phi^{\prime}(a_1^-)-H = -\left(\phi^{\prime}(a_n^+)-H\right),$$ 
but this equality is 
false for a non centered unit. It is possible to choose a correction $H_\nu$ to 
the magnetic field $H$ in order to obtain the equality:
\begin{eqnarray}
\phi^{\prime}(a_1^-) - H + H_\nu & = & -\left(\phi^{\prime}(a_n^+)-H + 
H_\nu \right),\nonumber\\
-\nu j a_1 + (1-\nu) {\gamma \over 2} + H_\nu & = & 
-\left[ \nu j (l-a_n) - (1-\nu) {\gamma \over 2} + H_\nu \right], \nonumber\\
\label{magn.shift}
H_\nu & = & \nu j \left(\frac{l_b-l}{2}+a_1\right).
\end{eqnarray}
Let us consider two arrays, $1$ with a centered junction unit and $2$ with 
the same junction unit but non centered.

\begin{proposition}[{\rm Magnetic shift}]
Let $(H,\gamma_{max})$ be the coordinates of a point of the
$\gamma_{max}(H)$ curve for the circuit $1$. Then $(H+H_\nu,\gamma_{max})$ is a 
point of the curve for the circuit $2$.
\end{proposition}

\begin{figure}
\centerline{ \epsfig{file=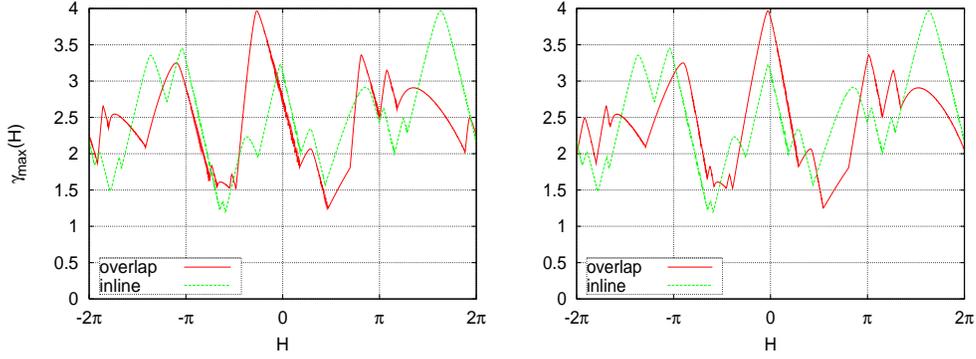,height=14 cm,width=5 cm,angle=270}}
\caption{Plot of $\gamma_{max}(H)$ for a four
junctions device ($l=10$, $d_i=1$) such $l_1=1.5$, $l_2=2.5$, $l_3=2$.
The left panel shows a non centered junction unit with $a_1=0.1$ and the 
right panel a centered unit with $a_1=2$. Notice the current dependent
shift (\ref{magn.shift}) for the overlap solution as one goes from a centered junction unit (right
panel) to an off-centered junction unit (right panel). The junction unit 
was moved to the left.}
\label{f3} \end{figure}

So, moving a junction unit translates the $\gamma_{max}$ curve by
$\nu j a_1$. Fig.~\ref{f3} shows a $\gamma_{max}(H)$ for a four
junctions device with a non centered junction unit in the left panel
and a centered junction unit in the right panel. Both inline and
overlap current feeds are presented. Notice the unchanged behavior for the inline
current feed and the effect of $H_{\nu}$ $(= - 4.1 \gamma / 10)$ from
(\ref{magn.shift}) in the overlap case.

{\em Proof}. Let $\phi_{\{H,\gamma,b_1\}}$ be a solution for an array
$A_1\equiv\{a_1,...,a_n\}$ with a centered 
junction unit with $\gamma$ and $H$ given. 
Consider another array $A_2$ with the same junction unit moved by $s$,
$A_2\equiv \{a_1+s,...,a_n+s\} \equiv A_1+s$, the coefficients $d_1,...,d_n$ 
being equal 
for the two circuits. From the solution $\phi_{\{H,\gamma,b_1\}}$ for $A_1$
we can deduce a solution
$\psi_{\{H+H_\nu,\gamma,b_1+s\}}$ for $A_2$. From (\ref{magn.shift}) we have: 
$$\psi^{\prime}_{\{H+H_\nu,\gamma,b_1+s\}}(a_1^- +s)=
\phi^{\prime}_{\{H,j,b_1\}}(a_1^-)$$
Taking, 
$$\psi_{\{H+H_\nu,\gamma,b_1+s\}}(a_1^- +s)=\phi_{\{H,\gamma,b_1\}}(a_1^-)$$
and from the unicity of the solution, we obtain $\phi \equiv \psi$ in the two
junction units.
Thus, if $\phi$ is a solution for $\{H,\gamma \}$ given for $A_1$, 
then $\psi$ is a solution for $\{H+H_\nu,\gamma \}$ for $A_2$,
and vice versa. 

Let $\gamma_{max,1}$ and $\gamma_{max,2}$ be the $\gamma_{max}$ curves for
the arrays $A_1$ and $A_2$. From the solutions obtained for $A_1$
we build solutions for $A_2$. As a consequence,
$\gamma_{max,1}(H+H_\nu)
\leq \gamma_{max,2}(H)$. On the other side, from solutions of $A_2$
we build solutions for $A_1$, then
$\gamma_{max,1}(H+H_\nu)\geq \gamma_{max,2}(H)$. So, we obtain the equality:
$$\gamma_{max,1}(H+H_\nu)=\gamma_{max,2}(H)~.$$\endproof \\
Notice that this equality is independent of the number of junctions.

\subsection{Comparison between inline and overlap current feeds}

We now compare the $\gamma_{max}$ curves for inline and overlap current feed.
For one junction, the problem can be solved exactly
using polynomials by parts (see remark \ref{rem1}). We obtain,
$\gamma_{max}(H)=d_1$, for all $\nu$.
For two junctions there is the possibility of $d_1 \neq d_2$ and this
will change $\gamma_{max}(H)$ qualitatively. Let us study the phase 
difference between two junctions. 
We use remark \ref{rem1} and the boundary conditions to get
\begin{eqnarray}
\phi_2-\phi_1 & = & -\frac{\nu j}{2}(a_2-a_1)^2 + \left(P^{\prime}(a_1)
+d_1\sin(\phi_1)\right)(a_2-a_1)~, \nonumber \\[-1.5ex]
\label{shift2j} \\[-1.5ex]
\frac{\phi_2-\phi_1}{a_2-a_1}& = & - \nu j \frac{a_2+a_1}{2} + H -
\frac{1-\nu}{2}\gamma + d_1 \sin(\phi_1)~. \nonumber
\end{eqnarray}
If $(a_2+a_1)/2=l/2$ (the junction unit is also centered), as
$\gamma = jl$, (\ref{shift2j}) becomes:
\begin{equation} \label{2jj inli}
\frac{\phi_2-\phi_1}{a_2-a_1} =   H - \frac{\gamma}{2}
+ d_1 \sin(\phi_1)~.
\end{equation}
Note that we can obtain (\ref{2jj inli}) from (\ref{shift2j})
wiçth $\nu=0$. We have shown that,
\begin{proposition}[{\rm Equivalence of all current feeds for a centered SQUID.}]
\label{pro2} For a centered two junctions device, all current feeds
give the same $\gamma_{max}$ curve.
\end{proposition}

\begin{figure}
\centerline{\epsfig{file=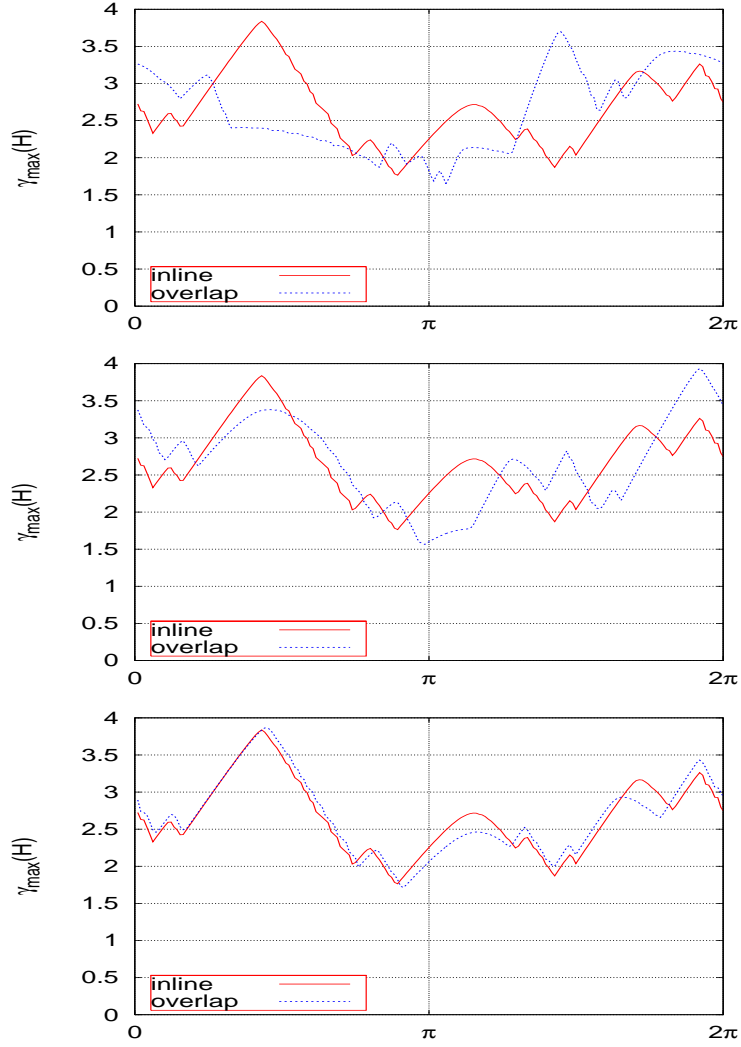,height=15 cm,width=10 cm,angle=0}}
\caption{Plot of $\gamma_{max}(H)$ for the same centered junction unit 
$l_1 = 1, l_2=4,l_3 = 3,~~d_1=d_2=d_3=d_4=1$ and different lengths $l$
of the microstrip, from top to bottom $l=8,16$ and 64. Notice
how the overlap solution tends to the inline solution as one
increases $l$.}
\label{f4} \end{figure}

For an inline current feed, $\nu=0$ so that the phase difference 
$\phi_2-\phi_1$ is independent of the position of  
the junction unit. This is not true for the overlap feed, 
where moving the junction unit causes a "magnetic shift" 
as seen above in $H_\nu$ equation (\ref{magn.shift}). 
When the number of the junctions $n\geq 3$, the $\gamma_{max}$ curve 
depends on $\nu$. The effect of the moving the junction unit on the 
$\gamma_{max}$ curve was shown above. So, we can reduce the study to 
a centered junction unit. In this case, we have $a_1=(l-l_b)/2$, and
\begin{eqnarray}\label{phia1inline}
\phi^{\prime}(a_1^-) & = & \phi^{\prime}(0) + \int_0^{a_1^-} 
-\nu \frac{\gamma}{l}dx, \nonumber \\
& = &H - (1-\nu) \frac{\gamma}{2} - \nu \frac{\gamma}{2} + \frac{\nu l_b}{l}
\frac{\gamma}{2} = H-\left(1-\frac{\nu l_b}{l}\right)\frac{\gamma}{2}, \\
\phi^{\prime}(a_n^+) & = & H +  \left(1-\frac{\nu l_b}{l}\right)\frac{\gamma}{2},
\nonumber
\end{eqnarray}
with $l_b=a_n-a_1$. We can write $\nu j=(\nu l_b/l)(\gamma/l_b)$, and 
$\nu l_b/l=\mu$. So, equation (\ref{e5.1}) is equivalent to the system:
\begin{equation}\label{e5.l_b}
-\phi^{\prime\prime}(x)+\sum_{i=1}^n d_i \delta(x-a_i)\sin(\phi) = \mu
\frac{\gamma}{l_b}~,
\end{equation}
with, $$ \phi^{\prime}(a_1^-) = H-(1-\mu)\gamma/2,~~
\phi^{\prime}(a_n^+) = H+(1-\mu)\gamma/2.$$
As $0\leq \nu \leq 1$, $0 \leq \mu \leq l_b/l$. 
Note that $l_b$ can be considered as the reference length of the device.
Also note that if $l \rightarrow +\infty$ then $\mu\rightarrow 0$ and
the equation (\ref{e5.l_b}) and boundary conditions
tend to the situation of inline current feed.
Fig.~\ref{f4}, illustrates this convergence when we increase the microstrip length 
$l$ for a centered junction unit. 
Notice that the solution for the inline feed
is not modified by the variation of length. With $l=8$, we have a maximum difference
between the solutions for the overlap and inline current feeds.
As $l$ increases the solution for overlap current feed tend to the solution
for inline current feed.
We prove this in the appendix 
'Convergence of the solution for a large length $l$'.

{\bf Conclusion:} In the appendix, we show that when $\nu \gamma/l$ tends to $0$
the solution tends to the one for inline current feed. We can get this
by increasing the length $l$ or shrinking the junction area.
(see in appendix 'Convergence by the coefficient $d_i$.'). We have 
three parameters: $\nu$, $l$ and $\gamma_{max}$.
$l$ is determined by the circuit. $\nu$ comes from the $2D$ model, 
it depends on the width of circuit. The third parameter, can be 
bounded from above: $0 \leq \gamma \leq \sum_i d_i$. We will see in the next 
section, what is the limit of $\gamma_{max}(H)$ for
inline and overlap feeds when $d_i$ are small.

\subsection{The relation between inline and magnetic approximation}

The size of the junctions $w_i<1<w$ so that $d_i <<1 $ therefore the
jump of the gradient of the phase across the
junctions can be neglected.  This is the magnetic approximation
where only $H$ fixes the phase gradient.
In the previous section, we have shown 
that the solution for inline and overlap current feeds converge to 
the same $\gamma_{max}(H)$ curve for small $d_i$. We will show that 
this limit is the magnetic approximation.

\begin{figure}
\centerline{\epsfig{file=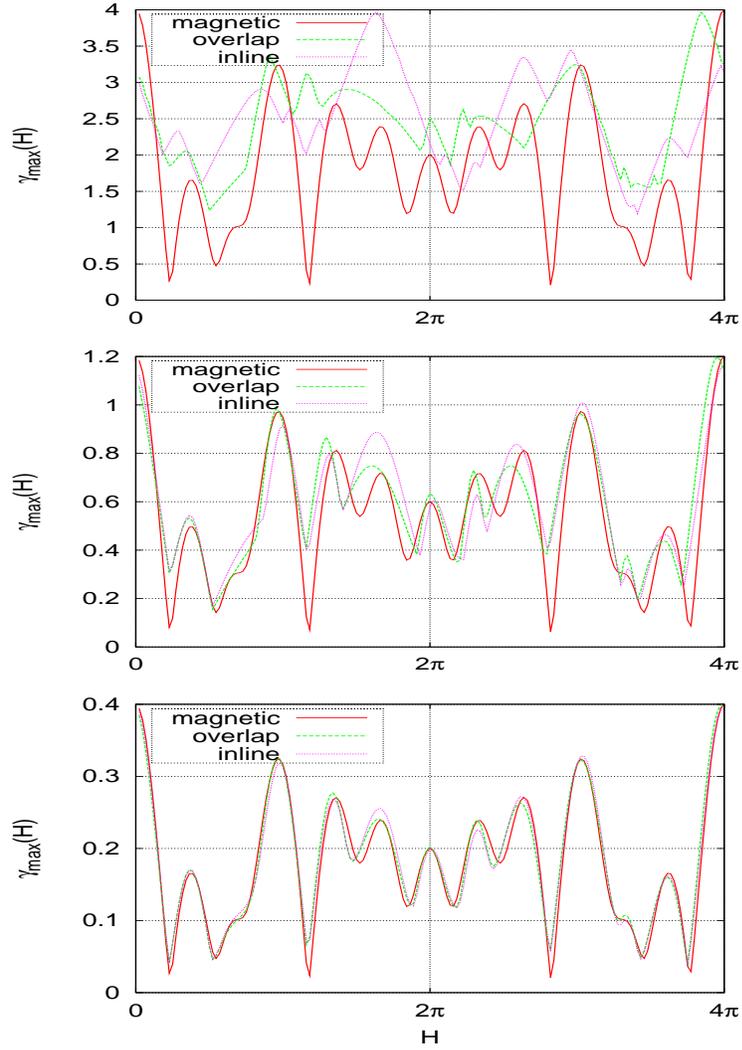,height=15 cm,width=10 cm,angle=0}}
\caption{Plot of $\gamma_{max}(H)$ for the same junction unit and different
coefficients $d_i$ which are all equal, from top to bottom $d_i=1,0.3$ and $0.1$. The 
distances between the junctions are
$l_1=1.5$, $l_2=2.5$, $l_3=2$, $l=10$.}
\label{f5} \end{figure}

Since $\left[\phi^{\prime}\right]_{a_i^-}^{a_i^+}=d_i \sin(\phi_i)$ 
(remark \ref{rem2}) and $j \leq \sum_i d_i/l$, then for small $d_i$, 
then $\phi$ tends to the linear function $\phi(x) = Hx + c$. 
This magnetic approximation seems crude but 
we show that it approaches the solution for inline feed,
see the appendix "Inline - magnetic convergence". 
There we bound the difference between the $\gamma$ curves for the
inline feed and the magnetic approximation.
Fig.~\ref{f5} illustrates this convergence as $d_i$ decreases.

This approximation gives very good results because we work on very 
small junctions and the corresponding $d_i\approx 10^{-2}$ (compared with 
the values taken in Fig.~\ref{f5}). 

\subsection{Magnetic approximation}

The magnetic approximation is very interesting because it gives an 
analytic expression of $\gamma_{max}(H)$ and is independent 
of the value of the current and of the scale of the circuit. Here we 
consider that $\phi(x)=Hx+c$ and from (\ref{rem3}) 
$$\gamma=\sum_{i=1}^n d_i \sin(Ha_i+c)~.$$ 
Notice that $c$ is the only parameter which can be adjusted to
reach the maximum.

To find the $\gamma_{max}(H)$ curve of the magnetic approximation, we 
take the derivative 
\begin{equation}\label{e3.3.1}
\frac{\partial \gamma}{\partial c} = -\sin(c)\left(\sum_{i=1}^n d_i\sin(Ha_i)\right)+
\cos(c)\left(\sum_{i=1}^n d_i\cos(Ha_i)\right).
\end{equation}
The values of $c$ canceling $\partial \gamma / \partial c$ are
\begin{equation}\label{cmax}
c_{max}(H) = \arctan\left(\frac{\sum_{i=1}^n d_i \cos(Ha_i)}{\sum_{i=1}^n d_i 
\sin(Ha_i)}\right)~,
\end{equation}
and using (\ref{e3.3.1}), we have the solution:
\begin{equation}\label{feyn}
\gamma_{max}(H) = \left| \sum_{i=1}^n d_i \sin(Ha_i + c_{max}(H)) \right|.
\end{equation}
This $\gamma_{max}$ curve is a function of $H$. A similar expression
was given by Miller et al \cite{mghg} for homogeneous arrays. Here we generalize 
this approach to nonhomogeneous arrays and justify it rigorously.

{\bf Remark:} \\

If $d_i=d$, $\forall i \in \{1,...,n\}$, we can simplify:
$$c_{max} = 
 \arctan\left(\frac{\sum_{i=1}^n \cos(Ha_i)}{\sum_{i=1}^n \sin(Ha_i)}\right)$$
In the same way,
$$\gamma_{max}=d\left|\cos(c_{max})\left(\sum_{i=1}^n \sin(Ha_i)\right) + \sin(c_{max})
\left(\sum_{i=1}^n \cos(Ha_i)\right)\right|.$$
Here changing the value of $d$ will change linearly the 
amplitude of $\gamma_{max}$ curve, this is not the case
for the solutions of the boundary value problem (\ref{e3.1}).
We can notice too, that the $\gamma_{max}(H)$ obtained from this
approximation is invariant 
by the transformation, \\
$\forall t \in \Re,~\left\{\begin{array}{lcl}
a_i &\rightarrow & ta_i \\
H &\rightarrow & \frac{1}{t} H
\end{array}\right.$. \\

We will show in the next sections that when $d_i<<1$, (\ref{feyn}) provides
a good estimate of the $\gamma_{max}(H)$ curve of a circuit. In addition,
from its invariant properties we can compare different models
and estimate the parameters of the circuit. 
It is a good approximation for the physical device. 
A cooperation has begun with the LERMA at the Observatoire
de Paris to match theory and design for this type of circuit with specific
properties\cite{sbcld}. 

\section{Numerical solutions}

We used two different methods, a stepping in the $(H,\gamma)$ plane using 
a Newton iteration and what we call the
method of implicit curves to find the maximal current of Eq.(\ref{e5.1}) for $H$ given. 

\subsection{Newton's method}

We start from the system of nonlinear transcendental equations (\ref{system})
which is written for $n=5$.
Introducing the vector $X=(\phi_1,\phi_2,...\phi_n)$, (\ref{system}) 
can be written as $F(X)=0$ where $F$ is a nonlinear map from $R^n$ 
to $R^n$. To 
solve numerically this equation, we use the Newton method.
$$X_{k+1}=X_k-(\nabla F(X_k))^{-1} F(X_k),$$
where $\nabla F(X_k)$ is the gradient of $F$ evaluated at $X=X_k$.
A first problem is to choose the initial vector $X_0$. For that consider $H = 0$, 
there we expect a 
solution such that $\gamma\approx \sum_i^n d_i$ consequently 
$\phi_i\approx \pi/2\left[2\pi\right]$. 
We have our initial vector. After finding the solution, we step in $H$ and take 
as initial guess, the previous solution found, which for a small step in magnetic
field is assumed to be close to the one we are looking for. 
By this way, we obtain a solution with a magnetic field $H+ dH$ and a current $\gamma$.
We can then increase $\gamma$ until the method does not converge and this gives
the maximum current $\gamma_{\rm max}(H + dH)$ for increasing $H$.  
Similarly we can compute $\gamma_{max}(H)$ by starting with a large magnetic field and
decrease $H$ to 0. This curve will in general be different from the one obtained
when increasing $H$ due to hysteresis. The two curves need to be overlapped
to see where is $\gamma_{max}(H)$.
So, we introduce another method to be sure to obtain directly $\gamma_{max}$ curve.

\subsection{Implicit curves method}

The polynomials (\ref{rec1}) and (\ref{recn}) establish the existence and value
of $\phi$ at the junctions. This function should satisfy the 
boundary conditions. The first one
$$\phi^{\prime}(0)=P_1^{\prime}(0)=H-(1-\nu)\gamma/2$$
is true by construction, the second (for $n$ junction circuit) is:
\begin{equation}\label{rbc}
P^{\prime}_{n+1}(l)=H +(1-\nu)\frac{\gamma}{2},
\end{equation}
is true only for the solutions of Eq.(\ref{e5.1}). As we have remarked in the
section "Polynomial by part", $\phi$ is entirely determined by $\phi_1$, 
$\gamma$ and $H$. For $H$ given, 
the solutions of Eq.(\ref{rbc}) define a relation between $\phi_1$ and $\gamma$.
So, the maximal current solution depends on $\phi_1$ and $\gamma$ and Eq.(\ref{rbc})
is the constraint it should satisfy. As the solutions $\phi$ are defined 
modulo $2\pi$, see (\ref{rem1}), we can assume $\phi_1\in[-\pi,\pi]$. On the other 
hand, because of (\ref{rem3}), $\gamma \in \left[0,\sum_i d_i\right]$.
To solve this problem with Maple\cite{maple}, we plot the implicit function (the constraint) of the 
two variables $\phi_1$ and $\gamma$ with $H$ and $\nu$ fixed, defined by
\begin{equation}\label{impli}
\left .P^{\prime}_{n+1}\right|_{x=l}
(\phi_1,\gamma,\nu,H)-H-\frac{1-\nu}{2}\gamma=0,
\end{equation}
with $(\phi_1,\gamma)\in[-\pi,\pi]\times\left [ 0,\sum_i d_i \right ]$.
The program searches, in an exhaustive way, the biggest value of 
$\gamma$ of this implicit curve. Incrementing $H$, we obtain the relation
$\gamma_{max}(H)$.
We give an expression of $P^{\prime}_{n+1}$ for two and three junctions, in
the Appendix: 'Implicit curves'.

\begin{figure}
\centerline{\epsfig{file=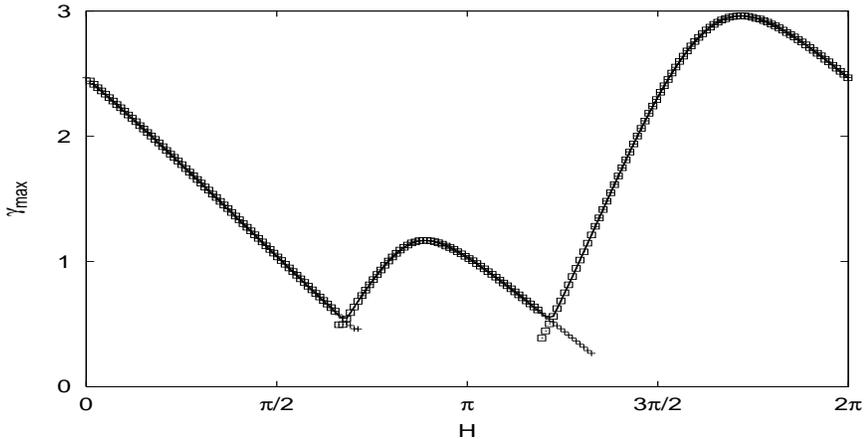,height=6 cm,width=12 cm,angle=0}}
\caption{Comparison between the Newton method and the implicit curve
method for the $\gamma_{max}$ curve for a three junction unit
$a_1=1,~a_2=2,~a_3=3$,$d_1=d_2=d_3=1,~\nu=1$ and $l=10$. The squares (resp. the $+$)
symbols correspond to the Newton results for decreasing (resp. increasing)
$H$ and the continuous line corresponds to the results of the implicit
curve method.}
\label{f5a} \end{figure}

Compared to the Newton method detailed in the previous section,
this method has the advantage to converge to a global maximum $\gamma_{max}$, 
as long as we give enough points to plot the implicit curve. Fig. \ref{f5a}
compares $\gamma_{max}(H)$ using the two methods for a three junction
unit. The solution given by the implicit curve method is in continuous line
and superposes exactly with the other two plots. With the 
Newton method we can get trapped in local maxima while
the implicit curve method always gives the global maximum. On the other hand
the Newton method is much faster.

\section{Two junctions}

We have seen two methods to solve the problem numerically and established
general properties.
Now let us use these results for an array with a few junctions.

\subsection{Same junction strength ($d=d_1=d_2$)}

\begin{figure}
\centerline{\epsfig{file=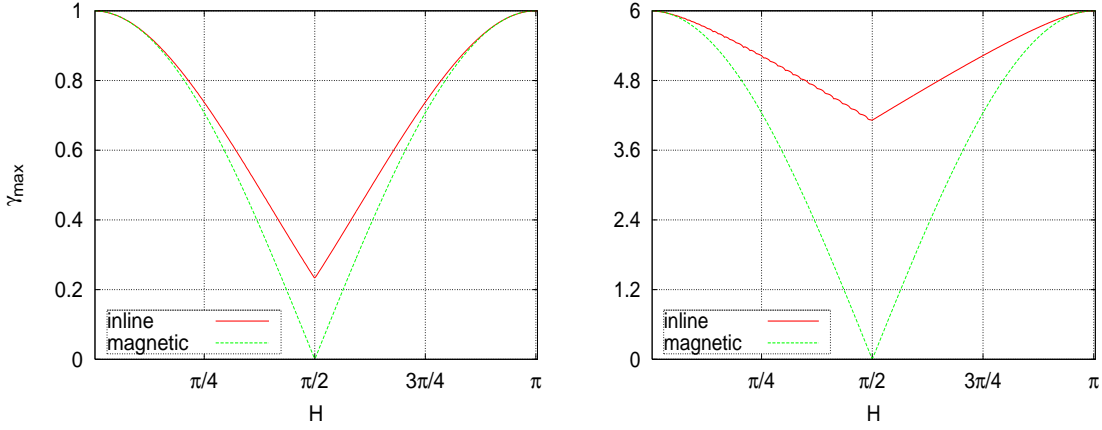,height=16 cm,width=6 cm,angle=270}}
\caption{Plot of the $\gamma_{max}$ curve for a two junction unit 
such that $l_1=2$. In the left panel, $d_1=d_2=0.5$ while on the right panel $d_1=d_2=3$. }
\label{f6} \end{figure}

\begin{figure}
\centerline{\epsfig{file=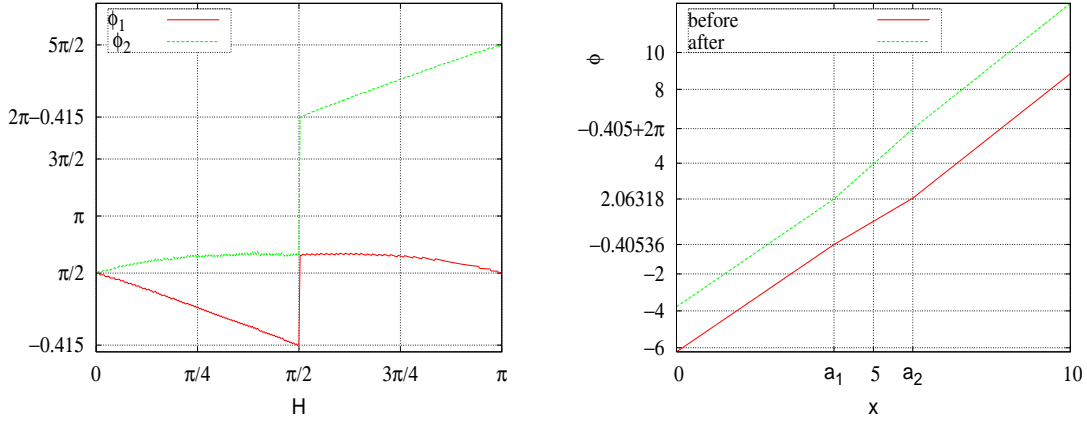,height=16 cm,width=6 cm,angle=270}}
\caption{Left panel: plot of the phases $\phi_1$ and $\phi_2$ as a function of the 
magnetic field $H$
for the same junction unit as the one shown in the left panel of Fig.~\ref{f6}
On the right panel, we plot $\phi(x)$ for the same device for 
$H < \pi /(a_2-a_1)$ (continuous line) and $H> \pi /(a_2-a_1)$ (dashed line).}
\label{f7} \end{figure}

In Fig.~\ref{f6}, we plot in the left panel $\gamma_{max}(H)$ of a two junction
unit.
We find the expected periodicity $H_p=2\pi /\left(a_2-a_1\right)$, with a maximum 
for $H=0$ in the inline case ($\nu=0$). For the overlap feed, we have exactly
the inline curve plus a magnetic shift. 
Notice that for the inline feed the amplitude of the $\gamma_{max}$ curve is not proportional
to $d_i$, contrary to the magnetic approximation.
The larger the $d_i$ the further away the $\gamma_{max}(H)$ curves are from the
ones given by the magnetic approximation. This is expected because the magnetic approximation
neglects the effect of $d_i$ on the phase.
In this section, to simplify the discussion we will restrict ourselves to the inline
current feed. However the results will be valid for the general case.
The maximum of $\gamma_{max}$ corresponding to $H=0~ mod. H_p$ is the only case where
$(\phi_2-\phi_1)/(a_2-a_1)=H$. On the other hand, by construction, in the magnetic approximation 
$(\phi_2-\phi_1)/(a_2-a_1)=H$ for all $H$. In the general case,
the closer $H$ is to $\pi /(a_2-a_1)$, the further $(\phi_2-\phi_1)/(a_2-a_1)$ is
from $H$.
This can be seen in the right panel of Fig.~\ref{f7}. So, 
there will be more tunneling current in one junction than in the other.
This phenomenon increases as $H$
increases from 0 to $\pi /(a_2-a_1)$. For that value, we have two possible 
solutions for $\gamma_{max}$ as shown in the left panel of Fig.~\ref{f7} for $H=\pi/2$.

As the field crosses $\pi /(a_2-a_1)$ the two junctions behave in opposite fashion
as shown by the switch of the jumps in $\phi_x$ at the junctions, see right panel of Fig.~\ref{f7}.
These two solutions or reversing behavior of junction, cause a
jump in $\gamma_{max}^{\prime}(H)$. As long as the evolution of $\phi_1$ (or $\phi_2$)
is continuous there is no jump in $\gamma_{max}^{\prime}$.
To summarize, the smaller $d$ is, 
the closer $(\phi_2-\phi_1)/(a_2-a_1)$ is to $H$.
Another way of relaxing this constraint on $(\phi_2-\phi_1)/(a_2-a_1)$
for a constant $d$, is to separate the junctions and we can show that 
$l_b=a_2-a_1 \rightarrow +\infty$ then $\gamma_{max}(H) \rightarrow d_1+d_2$.

\subsection{Regularity of $\gamma_{max}(H)$}

\begin{figure}
\centerline{\epsfig{file=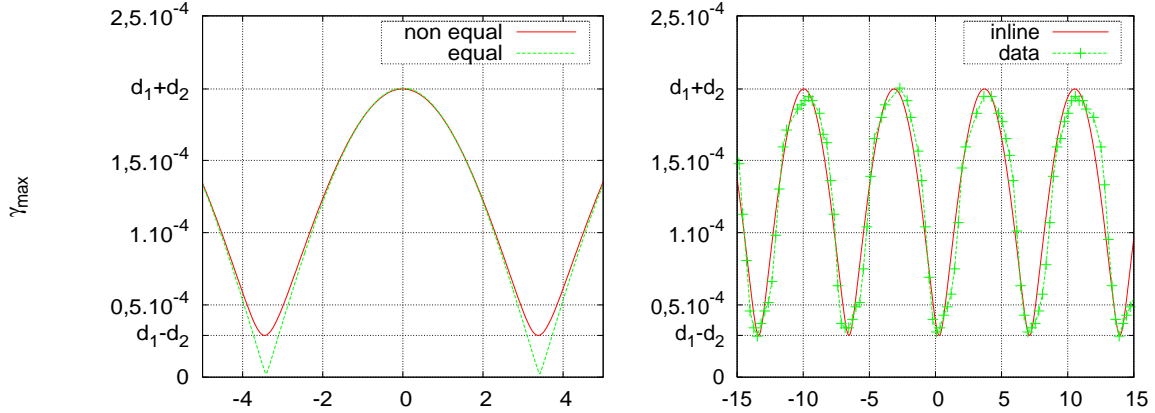,height=16 cm,width=6 cm,angle=270}}
\caption{Plot of $\gamma_{max}(H)$ curves for
two two junction units with inline current feed. On the left
panel we compare $\gamma_{max}(H)$ given by the model for the cases
$d_1=d_2$ and $d_1 \neq d_2$. The right panel shows the fit of the experimental
data from a two junction unit of the Observatory of Paris (Reproduced with permission
of Faouzi Boussaha and Morvan Salez).} 
\label{f8} \end{figure}

Junctions are never perfectly similar, small differences in their
areas or their critical currents will affect $\gamma_{max}$. In the 
left panel of Fig.~\ref{f8} showing $\gamma_{max}(H)$ for a two junction device
there is no discontinuity
of the slope of the curve $\gamma_{max}(H)$ labeled "non equal", 
$\partial \gamma_{max}/\partial H$ exists everywhere. In this case, 
the value of $\phi_1(H)$ and $\phi_2(H)$ associated to 
$\gamma_{max}(H)$ vary continuously.

To show this, consider a circuit such that $d_1>d_2$. Fixing 
$\phi_1=\pi/2$, implies that $d_1-d_2 \leq \gamma$ and
consequently $d_1-d_2 \leq \gamma_{max}$. 
From remark (\ref{rem3}) we have $\gamma_{max} \leq d_1+d_2$. Combining these 
inequalities we get $d_1-d_2 \leq \gamma_{max} \leq d_1+d_2$.
Now let us find the values of $H$ for which these bounds can be 
reached. From remark (\ref{rem3}) and equation (\ref{recn}) we have
\begin{eqnarray}\label{e6.1}
\gamma &=& d_1 \sin(\phi_1)+ d_2 \sin(\phi_2)~, \nonumber\\
\phi_2 &=& -\frac{\nu j}{2}l_1^2+ 
\left(H-\left(\nu a_1+\frac{1-\nu}{2}l\right)j+d_1\sin \phi_1 \right)l_1 
+\phi_1~, \end{eqnarray}
with $l_1=a_2-a_1$. By substituting the second equality into the first, 
and taking the derivative with respect to $\phi_1$, we obtain
$$\frac{\partial \gamma}{\partial \phi_1} = d_1 \cos \phi_1 + d_2
\left[\left(-\left(\nu \frac{a_2+a_1}{2l}+\frac{1-\nu}{2}\right)
\frac{\partial \gamma}{\partial \phi_1} +d_1
\cos\phi_1\right)l_1+1\right]\cos\phi_2,$$
Since we search for the maximum of $\gamma$, then
$\partial \gamma / \partial \phi_1 = 0$, so that
\begin{equation}\label{e6.2}
d_1 \cos \phi_1 = - d_2 \left(d_1 l_1\cos(\phi_1) + 1\right)\cos\phi_2~.
\end{equation}
When $\phi_1=\pi/2$, this condition gives $\phi_2 = \pi/2~ {\rm modulo}~~ \pi$. 
Now, inserting these solutions in (\ref{e6.1}), we obtain the values of 
$H$ for which these solutions are possible.
$$\begin{array}{|l|l|}
\hline 
\gamma_{max}(H) & H \\ \hline
d_1+d_2 & 2k\pi/l_1 +[\nu (a_1 +l_1/2)+(1-\nu)l/2](d_1+d_2)/l -d_1 \\ \hline
d_1-d_2 & (2k+1)\pi/l_1 +[\nu(a_1 + l_1/2)+(1-\nu)l/2 ](d_1-d_2)/l-d_1 \\ \hline
\end{array}$$
This enables from the curve $\gamma_{max}(H)$ to estimate $d_1$ and $d_2$.

We now proceed to give the condition between $d_1$ and $d_2$ 
such that the behavior of the $\gamma_{max}$ relation changes.
Since $\phi_2(H)$ varies continuously,
$\cos(\phi_2)$ takes all the values between $-1$ and $1$. 
We assume that $\forall \phi_1$, $d_1 l_1\cos(\phi_1) + 1\geq 0$
where $l_1=a_2-a_1$\footnote{for small junctions this is not a 
strong constraint, because
since $d_i=w_i^2/w<<1$, $w_i$, $w <<1$ and $l_i<<1$ are about $10^{-2}$.}.
We consider two cases:
\begin{remunerate} \item
$\cos\phi_1 \leq 0$: as $\cos\phi_1 \geq -1$ from (\ref{e6.2}) we obtain,
$$d_2\cos(\phi_2)\leq \frac{d_1}{1-d_1 l_1}~.$$
Since $\cos(\phi_2)$ must take all values between $-1$ and $1$ and $d_2>0$
\begin{equation}\label{limsup}
d_2 \leq \frac{d_1}{1-d_1 l_1}~.
\end{equation}
This is the maximal value than $d_2$ can take compared to $d_1$.
\item $\cos\phi_1 \geq 0$: as $\cos\phi_1 \leq 1$, for the same reason
we obtain:
\begin{equation}\label{liminf}
d_2 \geq \frac{d_1}{1+d_1 l_1}~.
\end{equation}
\end{remunerate}
To summarize $d\gamma_{max}(H)/dH $ does not vary continuously if 
\begin{equation}\label{dgdH}
{d_1  \over 1 -d_1 l_1}  \le d_2 \le {d_1  \over 1 + d_1 l_1}.\end{equation}

To illustrate this effect we consider the configuration of a
a microstrip with inline current feed with two Josephson junctions
built by Morvan Salez and Faouzi Boussaha at the Observatoire de Paris.
The results are shown in Fig.~\ref{f8}. 
The square junctions have an area of $w_j^2 \approx 1\mu m^2$, 
the Josephson length is $\lambda_J=5.6\mu m$ and $l_1=a_2-a_1=13 \mu m$ 
(using the junction centers). This gives  
$d_1=d_2\approx 0.0357$, $l_1 \approx2.32$ if the areas are equal. However
the experimental data does not go to 0 so that the junctions are probably slightly
different as expected from (\ref{limsup}) and (\ref{liminf}),
$$0.032969 \leq d_2\leq 0.038923~.$$
Only a $10\%$ difference in area is enough to give a regular
$\gamma_{max}(H)$. 
From the fit of the experimental data (right panel of Fig.~\ref{f8}) we can estimate
the areas of the junctions as $w_1^2 = 0.85255\mu m^2$ and $w_2^2 = 1.1417 \mu m^2$.

As we have seen in the previous section, when the $\gamma_{max}$ curve does not
show any spike, it is bounded by $d_1+d_2$ and $|d_1-d_2|$. From this
we can obtain the characteristics of the two junctions, their critical
current density and area except that 
we do not know which junction corresponds to $d_1$ and which to 
$d_2$. However if the $\gamma_{max}$ does not have any spikes then we
can give the exact area of the junctions assuming the critical density current
is known.

\section{Many junctions}

A two junctions circuit is a SQUID and shows a
simple $\gamma_{max}(H)$. To obtain specific properties for
advanced detectors, experimentalists make devices with more junctions.

\subsection{3 Josephson junctions}

When we add a new junction to a circuit with two junctions, new
oscillations appear on $\gamma_{max}(H)$. 
We cannot predict the amplitude of the oscillations but from
$l_1$ and $l_2$ we can estimate the number of oscillations in one 
period i.e. the interval $[0,H_{p}]$. 
\begin{figure}
\centerline{\epsfig{file=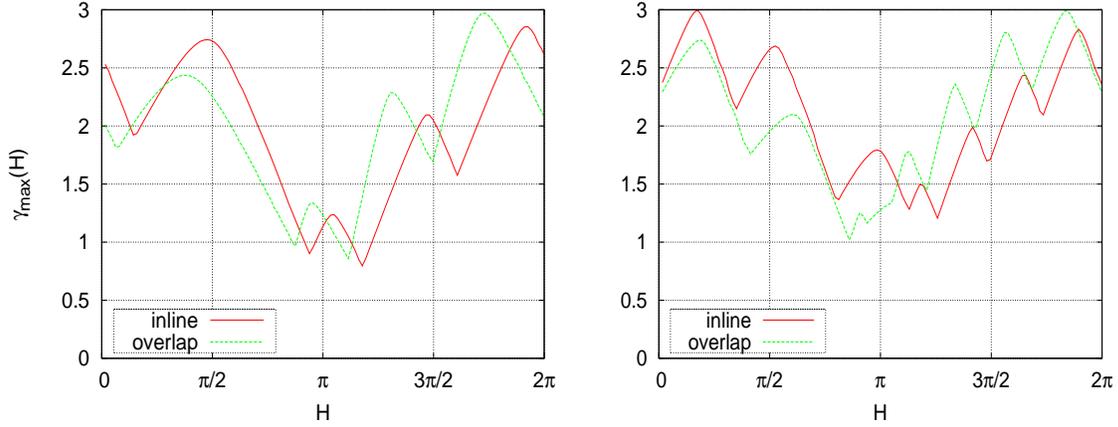,height=16 cm,width=6 cm,angle=270}}
\caption{Plot of $\gamma_{max}(H)$ curves for a two junction 
unit $a_1=1,~a_2=2$ ($l_1=1$) and a third junction placed at 
$a_3=5,~l_2=3$ (left panel) and $a_3=8,~l_2=6$ (right panel). All
the junctions have the coefficient $d_i=1$.}
\label{f9} \end{figure}
We intoduce the phase difference for $H=0$, $\Delta \phi_i 
\equiv \phi_i-\phi_1$.
Using Proposition \ref{pro1}, we can state that as $H$ goes from 0 to $H_p$,
$\phi_2-\phi_1$ goes from $\Delta \phi_2$ to $\Delta \phi_2 + 2 \pi l_1 /l_1=
\Delta \phi_2 + 2 \pi$. Similarly 
$\phi_3-\phi_1$ goes from $\Delta \phi_3$ to 
$\Delta \phi_3 + 2 \pi(l_2+l_1)/l_1=\Delta \phi_3 + 2\pi (l_2/l_1+1)  $ which
becomes $\Delta \phi_3 + 2\pi(k+1)$ if
the junctions are placed harmonically so that $l_2=kl_1$. In that case
we expect the $\gamma_{max}(H)$ curve to present $k+1$ bumps within one
period.
In Fig.~\ref{f9}, the junctions 
are placed in a harmonic way $a_3-a_2=k(a_2-a_1)$, where $k=3$ (left panel) and $k=6$ (right
panel). As expected we see the 4 intermediate "bumps" in the $\gamma_{max}(H)$ curve in the left
panel and 7 "bumps"in the curve of the right panel. 
We can see the periodicity
given by $H_{p}=2\pi/(a_2-a_1)\equiv 2\pi/ l_1$, which adds new oscillations. This
picture shows that the closer the third junction is to the junction unit
the fewer oscillations there are. Then the oscillations have a larger amplitude.
These estimations hold approximately in the case of an array with more junctions.

\begin{figure}
\centerline{\epsfig{file=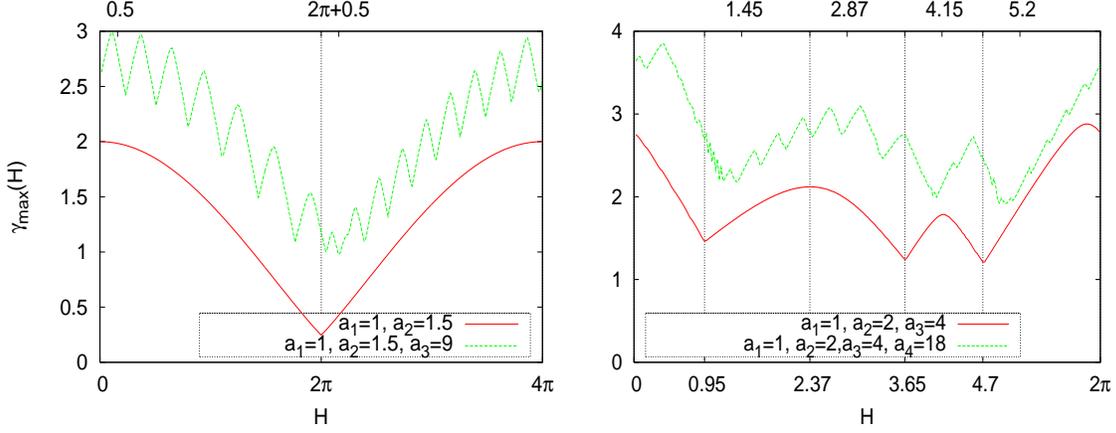,height=16 cm,width=6 cm,angle=270}}
\caption{Plot of $\gamma_{max}(H)$ showing the influence of a far away junction
on a junction unit. In continuous line we plot $\gamma_{max}(H)$ for the junction
unit only and in dashed line we plot $\gamma_{max}(H)$ for the junction unit
together with the far away junction. The left panel shows a two junction unit 
together with a third junction and the right panel a three junction unit
together with a four junction. The $H$ scale on the bottom of the graphs
indicates local minima or maxima of the junction unit curves and the
scale at the top shows these locations shifted by the quantity (\ref{n,n+1}).
the continuous line corresponds to the junction unit only
For all devices $d_i = 1$.}
\label{f10} \end{figure}
In other words, when $a_3-a_1$ is large as in 
the right panel of Fig.~\ref{f9} and the left panel of
Fig.\ref{f10}, the shape of $\gamma_{max}$ curve tends 
to the one for a two junctions circuit. We explain this below.

\subsection{Influence of a faraway single junction for the inline 
current feed}

In Fig.~\ref{f10}, for each panel, we plot a $\gamma_{max}$ curve, for
a $n$ junction unit, and another with the same junction unit plus a
far away junction. The $\gamma_{max}$ curve for $n+1$ junctions 
look like $n$ junctions curves to which a shift has been added. Let us
evaluate this shift.

Remark that for the junction $n$, using the
notations of Eq. (\ref{rec1}) and (\ref{recn}), we know that $\phi_{n+1}$ is
determined by $P_{n+1}$. If we increase $P_{n+1}^{\prime}$ of $\epsilon$, 
then $\phi_{n+1}$ increase of $\epsilon(a_{n+1}-a_n)$. 
Thus, a variation at $\phi_n$ of $\epsilon=\pm \pi/(a_{n+1}-a_n)$ 
is enough to obtain $\sin\phi_{n+1}=1$. The farther the last junction, the 
smaller 
$\epsilon$, and consequently this junction has the smallest action on the
junction unit. So, in the search of $\gamma_{max}$, the value of 
$\sin\phi_{n+1}$ is near
$1$. The $\gamma_{max}^{n+1}$ curve of a circuit with $n+1$ junctions is close
to $\gamma_{max}^{n}+d_{n+1}$ i.e. the curve for the $n$ junction circuit 
with $n$ junctions plus the maximal contribution of the last junction.

Now let us assume that $\sin\phi_{n+1}=1$. Let us recall the boundary
conditions of our inline current feed problem: 
$\left. \phi^{\prime}\right|_{\{0,l\}} = H \mp \gamma/2$. Therefore
the boundary conditions at the junction unit are $\phi^{\prime}(0) = H-\gamma/2$
and for $x$ such that $a_n<x<a_{n+1}$, $\phi^{\prime}(x) = H+\gamma/2-d_{n+1}$. 
As we have done in the section "magnetic shift", let $H^{\prime}=H-d_{n+1}/2$.
The previous boundary conditions become 
$$\left. \phi^{\prime}\right|_{\{0,a_n^+\}} = H^{\prime} \mp \frac{\gamma}{2}~.$$
We find the desired boundary values. Finally we obtain:
\begin{equation}\label{n,n+1}
\lim_{a_{n+1}-a_{n} \rightarrow +\infty} 
\gamma_{max}^{n+1}\left(H+\frac{d_{n+1}}{2}\right)
=\gamma_{max}^{n}(H)+d_{n+1}
\end{equation}
Fig.~\ref{f10} illustrates this convergence. 

This argument can not be extended simply 
to the overlap or general current feed for two reasons. 
First introducing or taking out the last junction $a_{n+1}$ induces
a variation of the magnetic shift $H_\nu$ given by (\ref{magn.shift}). We could
estimate it but we have the problem that the
curvature of $\phi$, for $n$ junctions device is $\nu j/2$ where $j$
depends on the number of the junctions. This will affect the shift
between the junctions and consequently the curve $\gamma_{\rm max}$.

However, numerical simulation show that Eq. (\ref{n,n+1}) remains 
a good approximation
for the general case (same order from inline) even with a small 
number of junctions. The general feed and inline feed problems 
coincide when $d_{n+1}/\sum_1^n d_i$ tends to zero. 
Going back to the physical device, this means that the
forces of the junctions are very small
$d_i \approx 10^{-2}$ and for these values
the inline and overlap results are practically indiscernible from
the magnetic approximation. Then (\ref{n,n+1}) can be used.

\subsection{A real device with 5 Josephson junctions}

\begin{figure}
\centerline{\epsfig{file=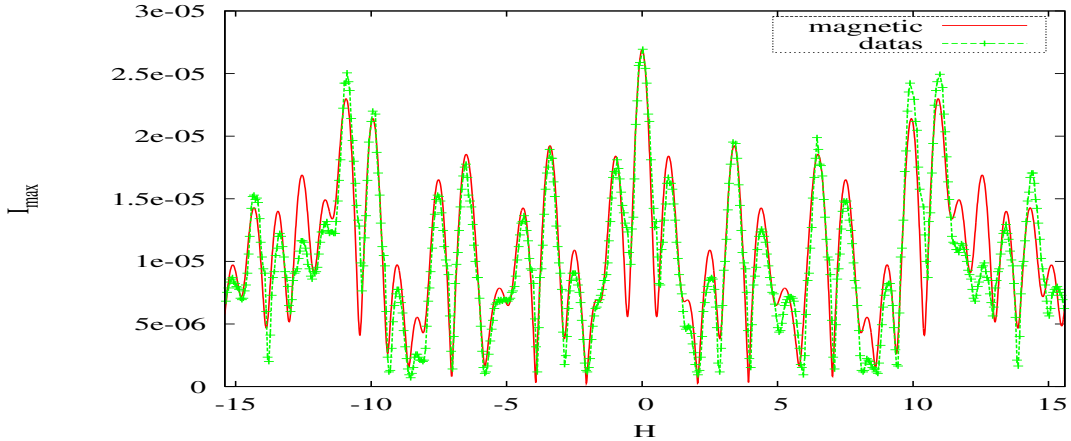,height=15 cm,width=6 cm,angle=270}}
\caption{Experimental $I_{\rm max}(H)$ for an array of five junctions in a
2D microstrip line built by M. Salez and F. Boussaha of the Observatory of
Paris (Reproduced with their permission). The measured data is 
presented by the $+$ symbols
and the magnetic approximation result is in continuous line.} 
\label{f11} \end{figure}

We have compared our theory to the experimental results for a device with 
two Josephson junctions. The same team at the 
Observatoire de Paris, has made a device with five junctions. Here the 
$\gamma_{max}$
curve obtained is totally different from the one for a simple SQUID. The parameters are
$l_1=20,~l_2=42,~l_3=12$ and $l_4=6$. Fig.~\ref{f11} shows 
the $\gamma_{max}$ curve where
the current and magnetic field have been scaled using approximately 
the same factors as for the SQUID of Fig. \ref{f7}. Our modeling approach also
gives excellent agreement for experimental uniform arrays of 5, 10 and 20 junctions.

\section{Conclusion}

We have analyzed mathematically a new continuous/discrete 
model for describing arrays of small Josephson
junctions. Compared to standard "lumped" approaches, we do not approximate
the equations, except for neglecting the phase variation in the junction. In
particular our approach preserves the matching at the interface.

We establish the periodicity of the $\gamma_{max}(H)$ curve, show how
it depends on the position of the array with respect to the microstrip. This
is particularly interesting to estimate the proportion of inline current feed
versus overlap feed. We show how separating a junction from
an array will influence $\gamma_{max}(H)$.

We introduce a numerical method for estimating $\gamma_{max}(H)$ which is
more reliable than the standard Newton method used up to now.

The relative simplicity of the model allows in depth analysis that
is out of reach for the 2D model. In particular 
we show that solutions for general current feed tend to the solutions
of inline feed when $\nu j / l \rightarrow 0$. All models reduce to what
we call the magnetic approximation for small $d_i$.

Our global model gives a very good agreement with experimental
curves obtained for arrays of up to five junctions. The simplicity of the
magnetic approximation allows to address the Inverse problem of determining
features of the array from $\gamma_{max}(H)$.

{\bf Acknowledgements}

J.G.C. and L. L. thank Faouzi Boussaha and Morvan Salez for helpful discussions
and for their experimental results. The authors thank Yuri Gaididei for
useful suggestions. The computations were done at the Centre de Ressources
Informatiques de Haute-Normandie (CRIHAN).

\section{Appendix}

\subsection{Implicit curves}

In this part, we give an example, of $P^{\prime}_{n+1}(x)$ for systems
with three junctions. We denote:\\
$\left\{ \begin{array}{rcl}
\sin_1 & = & \sin(\phi_1), \\ 
C_1 & = &\left(d_1\sin(\phi_1)-\frac{\nu\gamma a_1}{l}
+H-\left( 1-\nu \right) \frac{\gamma}{2}\right)(a_2-a_1)+\phi_1, \\
D_j & = & \frac {\nu\gamma(a_{j+1}-a_{j})^2}{2l}.
\end{array}\right.$\\
Then equations (\ref{rec1}) and (\ref{recn}) give
\begin{equation}\label{polybp2}
P_3^{\prime}(x) = -\frac{\nu\gamma x}{l} + 
d_2\sin(-D_1+C_1) + d_1\sin_1 + H-( 1-\nu) \frac{\gamma}{2}.
\end{equation}
\begin{eqnarray}\label{polybp3}
P_4^{\prime}(x)&=&-\frac{\nu \gamma x }{l} + d_3\sin \left[- D_2 + 
\left\{-d_2 \sin(D_1-C_1) -\frac{\nu \gamma a_2}{l} \right. \right. \nonumber \\ 
&& \left. \left. + d_1 \sin_1 + H-(1-\nu)\frac{\gamma}{2}
\right\}(a_3-a_2)-D_1+C_1\right] \nonumber \\ 
&& + d_2\sin(-D_1+C_1) + d_ 1\sin_1 +H-( 1-\nu)\frac{\gamma}{2}.
\end{eqnarray}
This example shows that $P_k^{\prime}(x)$ is $C^{\infty}$ in 
the variables $(\gamma,\phi_1,\nu,H,x)$. In particular $P_n^{\prime}(l)$
is $C^{\infty}$ in the variables $(\gamma,\phi_1,\nu,H)$.

\subsection{The current feed factor $\nu$: analytical estimates}

Equation (\ref{e5.l_b}) shows that we tend to an inline current feed when
$l$ is large. However we should show that the $\gamma_{max}$ curve
tends to the one for the inline feed.
\begin{lemma}[{\rm Solution}]
\label{solution}
For all $\phi_1$ and $H$, there exists a $\gamma$ such that 
equation (\ref{e5.1}) has a solution. 
\end{lemma}

{\em Proof}. As we have seen in the section "Implicit curves method", it is sufficient to 
solve equation (\ref{rbc}): $P^{\prime}_{n+1}(l)=H+(1-\nu)\gamma/2$, 
to find a solution. Let us fix a value for $\phi_1$ with $\nu$, $H$, $l$ given. 
If, $\gamma < -\sum_{i=1}^n d_i$, then 
$P^{\prime}_{n+1}(l) < H +(1-\nu)\gamma/2$. 
Conversely when $\gamma > \sum_{i=1}^n d_i$, we obtain 
$P^{\prime}_{n+1}(l) > H +(1-\nu)\gamma/2$. \\
But by construction, $P^{\prime}_{n+1}(l)$ is a function 
continuous in all its variables, in particular $\gamma$. Thus we have
at least one value of $\gamma$ in $[-\sum_{i=1}^n d_i,\sum_{i=1}^n d_i]$, such that
$P^{\prime}_{n+1}(l) = H +(1-\nu)\gamma/2$ so that it is
a solution for that value of $\phi_1$.\endproof

\begin{figure}
\centerline{\epsfig{file=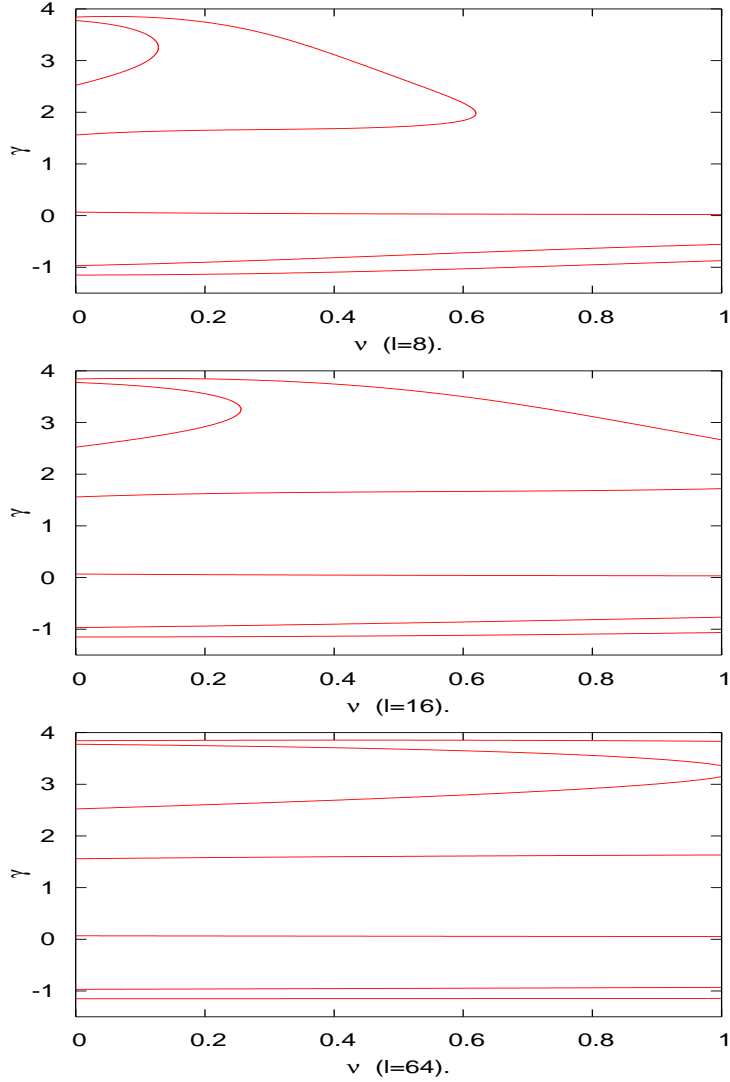,height=15 cm,width=10 cm,angle=0}}
\caption{Each panels corresponds to device of the panels of the Fig.~\ref{f4}.
We plot the implicit curve $\gamma(\nu)$ curve for $H=1.3617$, $\phi_1=1.3897$.
This coordinates give the maximum of the $\gamma_{max}(H)$ of inline of the top panels
of the Fig.~\ref{f4}. From top to bottom we increase the length of the device
and we notice the stretching of the $\gamma(\nu)$ curve (with the coefficient found
in equation (\ref{e5.l_b}): $l_b/l$).}
\label{f12} \end{figure}

We want to study the variation of $\gamma(H)$ versus the current feed $\nu$. At this point, we do not 
consider the $\gamma_{max}$ curve. Let us fix $\phi_1$. Using the previous property, we know
that there exists at least one solution of equation (\ref{e5.1}), and particularly
almost one $\gamma$. Without changing $\phi_1$ or $H$, we plot all the possible $\gamma$ 
versus $\nu$ in Fig.~\ref{f12}. We call this curve $\gamma(\nu)$ curve. 
To plot this $\gamma(\nu)$ curve, we use the same parameter as in Fig.~\ref{f4}, 
with $H=1.3617$ (see top panel, we choose this $H$ because there is a big 
difference between the solution for inline and overlap current feeds). We choose for $\phi_1$
the value found with Maple giving the maximum $\gamma_{max}$ for the inline feed. 
Fig.~\ref{f12}, top panel, for $\nu=0$, confirms the $\gamma_{max}$ value
found in Fig.~\ref{f4}. But for overlap the maximum current we can obtain is near $0$.
So, there is another value of $\phi_1$ for $\gamma_{max}$ of overlap current feed 
($\phi_1\approx0.252$).

Let us study $\gamma(\nu)$ curve. By definition, 
$\gamma = \sum_{i=1}^n d_i\sin(\phi_i)$. Let $\phi_1$ be a value such
$\partial \gamma/\partial \nu$ exists, then:
\begin{eqnarray}\label{dg/dn}
\frac{\partial \gamma}{\partial \nu} & = & \sum_{i=1}^{n} d_i 
\frac{\partial \phi_i}{\partial \nu} \cos(\phi_i),\nonumber \\
\left|\frac{\partial \gamma}{\partial \nu}\right|& \leq &
\sum_{i=1}^{n} d_i \left|\frac{\partial \phi_i}{\partial \nu}\right|.
\end{eqnarray}
With $\phi_i=\phi(a_i)$, and we note in the following $\phi_i^{\prime}=
\lim_{\epsilon\rightarrow 0} \phi^{\prime}(a_i-\epsilon)$ (the left derivative 
of $\phi$). Let us make some remarks: as $\phi_1$ is fixed, 
\begin{eqnarray*}
\left.\frac{\partial \phi_1}{\partial \nu}\right|_{\nu=0} & = & 0 \\
\frac{\partial \phi_1^{\prime}}{\partial \nu} & = & \frac{\partial}{\partial \nu}
\left\{H-\left(\frac{1}{2}-\frac{\nu l_b}{2l}\right)\frac{\gamma}{2}\right\}
=  -\left(\frac{1}{2}-\frac{\nu l_b}{2l}\right)\frac{\partial \gamma}{\partial \nu}+
\frac{l_b \gamma}{4l}, 
\end{eqnarray*}
using (\ref{rec1}) and (\ref{recn}) we can begin iteration,
\begin{eqnarray*}
\frac{\partial \phi_i}{\partial \nu} & = & - \left(\nu
\frac{\partial \gamma}{\partial \nu}+\gamma\right)
\frac{l_{i-1}^2}{2l} + l_{i-1} \frac{\partial \phi_{i-1}^{\prime}}{\partial \nu} 
+ \frac{\partial \phi_{i-1}}{\partial \nu}(d_{i-1} l_{i-1} \cos(\phi_{i-1})+1), \\
\frac{\partial \phi_i^{\prime}}{\partial \nu}& = & -\left(\nu
\frac{\partial \gamma}{\partial \nu}+\gamma\right)
\frac{l_{i-1}}{l} +d_{i-1}\frac{\partial \phi_{i-1}}{\partial \nu}
\cos(\phi_{i-1})+\frac{\partial \phi_{i-1}^{\prime}}{\partial \nu},
\end{eqnarray*}
with $l_i=a_{i+1}-a_i$. This last equation can be written, for $i\geq3$:
$$\left.\frac{\partial \phi_i^{\prime}}{\partial \nu}\right|_{\nu=0}=-\left(\nu
\frac{\partial \gamma}{\partial \nu}+\gamma\right)\frac{a_i-a_1}{l}
+\left.\frac{\partial \phi_1^{\prime}}{\partial \nu}\right|_{\nu=0}
+\sum_{k=2}^{k-1}d_k \frac{\partial \phi_k}{\partial \nu}\cos(\phi_k).$$
We obtain,
\begin{eqnarray*}
\frac{\partial \phi_{i+1}}{\partial \nu}& = &
-K_1^i \frac{\partial \gamma}{\partial \nu}
-K_2^i \gamma + \frac{\partial \phi_i}{\partial \nu}+
l_i\sum_{k=2}^i d_k \frac{\partial \phi_k}{\partial \nu}\cos(\phi_k),\\
K_1^i & = & l_i \left[\frac{\nu l_i}{2l}+\nu \frac{a_{i+1}-a_1}{l}
+\frac{1}{2}-\frac{\nu l_b}{2l}\right], \\
K_2^i & = & l_i\left[\frac{l_i}{2l}+\frac{a_{i+1}-a_1}{l}
-\frac{l_b}{2l}\right].
\end{eqnarray*}
Applying absolute values, we obtain:
\begin{equation}\label{majo1}
\left| \frac{\partial \phi_{i+1}}{\partial \nu}\right| \leq 
K_1^i \left|\frac{\partial \gamma}{\partial \nu}\right| + 
K_2^i \left|\gamma\right| + \left| \frac{\partial \phi_i}
{\partial \nu}\right| + l_i\sum_{k=2}^i d_k
\left| \frac{\partial \phi_{k}}{\partial \nu}\right|,
\end{equation}
We do not need to find exact expression of 
$|\partial \phi_{i+1}/\partial \nu|$, we know that it is
a linear combination of $|\partial \gamma/\partial \nu|$ and $|\gamma|$ and 
so is
$|\partial \phi_2/\partial \nu|$. Using (\ref{majo1}), we can show by
iteration that $|\partial \phi_i/\partial \nu|$ is a linear combination
of $|\partial \gamma/\partial \nu|$ and $|\gamma|$. Applying this last remark
to inequality (\ref{dg/dn}), we obtain that there exists two constants, 
$C_1$ and $C_2$ such
$$\left|\frac{\partial \gamma}{\partial \nu}\right| \leq C_1
\left|\frac{\partial \gamma}{\partial \nu}\right| + C_2 |\gamma|,$$
$C_1$ and $C_2$ are combination of $d_i$, $l_i$, $K_1^i$ and $K_2^i$. 
For $\nu$ and $d_i$ sufficiently small $|C_1|<1$, then
\begin{equation}\label{majo2}
\left|\frac{\partial \gamma}{\partial \nu}\right| \leq \frac{C_2}{1-C_1} 
|\gamma|.
\end{equation}
This last equation implies local continuity of the $\gamma$ curve as a function
of $\nu$. As we have seen in section "Comparison between inline 
and overlap" increasing $l$ is equivalent to decreasing the range of $\nu$ (given
by $\mu$). $\forall \epsilon$, $\exists L$/ $l\geq L$ $\Rightarrow$ $\mu \leq 
\epsilon$. This shows the convergence of $\gamma_{max}$ ($0\leq\nu\leq 1$) 
curve to inline current feed when $l\rightarrow + \infty$.

\subsection{Convergence by the junction coefficient $d_i$}

We want to show that the general case convergence to inline case, 
for small $d_i$. We have shown in the axiom of the previous appendix that 
for $H$ and $\phi_1$ given, we can find almost one solution, whatever $\nu$. This
show that for the same $\phi_1$, we can find a general and an inline solution.
Let us define:
\begin{remunerate}
\item $P^i_n(x)$, $\forall x \in ]a_n,a_{n+1}[$ a solution
of inline problem ($\nu=0$) of this circuit, $\gamma^i$ the maximal current
associated at the value $\phi_1$.

\item $P^g_n(x)$, $\forall x \in ]a_n,a_{n+1}[$ a general 
solution ($\nu\neq 0$, same $l$ and same junction unit), 
$\gamma^g$ the maximal current associated at the value $\phi_1$.

\item $\alpha_j$ and $\beta_j$ by:
$\left\{\begin{array}{rcl}
\alpha_j &=& P^{g^{ \prime}}_j(a_j) - P^{i^{\prime}}_j(a_j), \\
\beta_j &=& P^g_j(a_j) - P^i_j(a_j). \nonumber
\end{array}\right.$
\end{remunerate}
As $P^g_1(a_1)=\phi_1=P^i_1(a_,)$, we have $\beta_1=0$.
We can  calculate $\alpha_1$ using (\ref{phia1inline}),
$$\alpha_1=P^{g^{ \prime}}_1(a_1) - P^{i^{\prime}}_1(a_1)
=-\frac{\gamma^i-\gamma^g}{2}+\frac{\nu l_b}{2l}\gamma^g.$$
But $\gamma^i$ and $\gamma^g$ are positive, so
\begin{equation}\label{alpha1}
|\alpha_1| \leq \left(\frac{1}{2}+\frac{\nu l_b}{2l}\right)\sum_{i=1}^n d_i~.
\end{equation}
The aim of this following part is to bound $\beta_i$. 
We proceed by iteration. We recall that $l_k=a_{k+1}-a_k$. Using (\ref{recn}) 
we estimate $\beta_{k+1}$:
\begin{equation}\label{beta}
\beta_{k+1} = \frac{-\nu \gamma^g}{2 l}l_k^2+\left[
d_k (\sin(P^g_k(a_k))-\sin(P^i_k(a_k))) + \alpha_k \right]l_k+\beta_k.
\end{equation}
Let us focus on the sine terms,
\begin{eqnarray*}
\sin(P^g_k(a_k)) - \sin(P^i_k(a_k)) &=& 
\sin(P^i_k(a_k)+\beta_k) - \sin(P^i_k(a_k)),\\
&=& \sin(P^i_k(a_k))\left[\cos(\beta_k)-1\right]+\sin(\beta_k)\cos(P^i_k(a_k)),\\
&\leq& |\beta_k|^2 + |\beta_k|.
\end{eqnarray*}
We assume $\beta_k<<1$, thus we obtain the 
equivalences $\sin(\beta_k) \approx \beta_k$
and $\cos(\beta_k)-1 \approx -\beta_k^2$, but we cannot predict the sign of 
$ \sin(P^i_k(a_k))$ or $\cos(P^i_k(a_k))$. We neglect
$|\beta_k|^2$ compared to $\beta_k$. From (\ref{beta}), 
\begin{eqnarray*}\label{syslin}
|\beta_{n+1}| & \leq &  \left|\frac{\nu \gamma^g}{2l}l_n^2\right|
+ \left(d_n |\beta_n| + |\alpha_n|\right)l_n + |\beta_n|, \\
|\alpha_{n+1}| & \leq & \left|\frac{\nu \gamma^g}{l}l_n\right| 
+ d_n |\beta_n| + |\alpha_n|.
\end{eqnarray*}
Let us note $G =  \nu\sum_{i=1}^n d_i / l$, we obtain 
a simple system
\begin{equation}\label{zeta}
\zeta_{n+1} \leq M_n \zeta_n + G V_n~,
\end{equation}
with, $\zeta_n = \left(\begin{array}{c}
|\beta_n| \\ |\alpha_n|\end{array}\right)$, $M_n=\left(\begin{array}{cc}
d_n l_n+1 & l_n \\ d_n & 1
\end{array}\right)$ and $V_n = \left(\begin{array}{c}
l_n^2 /2 \\ l_n\end{array}\right)$.\\
So, we bound $|\beta_n|$ and $|\alpha_n|$, with $|\beta_1|$ and $|\alpha_1|$.
\begin{equation}\label{sysmat}
\zeta_n \leq M_{n-1}(...(M_2(M_1\zeta_1 + G V_1)+G V_2)...)+G V_{n-1}~.
\end{equation}
When $d_i \rightarrow 0$, 
\begin{remunerate}
\item $G\rightarrow 0$ then equation (\ref{sysmat}) tend to 
$\zeta_n \leq M_{n-1} \dots M_2 M_1\zeta_1.$
\item $M_k\rightarrow\left(\begin{array}{cc} 1 & l_k \\ 0 & 1\end{array}\right)$
then, $M_k \dots M_2 M_1 \rightarrow 
\left(\begin{array}{cc} 1 & \sum_{i=1}^{k-1} l_i \\ 0 & 1\end{array}\right).$
\end{remunerate}
From the two previous points, we obtain that 
$$|\beta_i|\leq |\beta_1|+|\alpha_1|(a_i-a_1)+O\left(\sum_{i=1}^n d_i\right)~.$$
Using (\ref{alpha1}), we have $|\alpha_1|(a_i-a_1)\leq \left(l_b/2+
\nu l_b^2/(2l)\right)\sum_{i=1}^n d_i$, and previous inequality 
become, $\forall i \in \{1,\dots,n\}$
\begin{equation}\label{majo3}
|\beta_i|\leq |\beta_1|+ O_1\left(\sum_{i=1}^n d_i\right)~.
\end{equation}
Remember that we have seen at the beginning that $\beta_1=0$, (\ref{majo3}) show
that $\gamma^g$ tend to $\gamma^i$. Since this convergence occurs independently 
of $\phi_1$, we obtain the convergence of the $\gamma_{max}$ curve.

\subsection{Inline - magnetic convergence}

We want to show in this part, the convergence of an inline solution to the 
magnetic approximation when $d_i<<1$. We already know that in this 
case the $\gamma_{max}$ curve for the general current feed and inline feed
tend to be equal. By this way, we show that for all $\nu$,
the $\gamma_{max}$  curve of Eq.(\ref{e5.1}) tends to the magnetic 
approximation when $d_i<<1$.

We know that the magnetic approximation is given by $f(x)=H x+c_{max}(H)$. Notice
that $c_{max}$ does not depend on the value of $\gamma$, see (\ref{cmax}). We 
are going to compare the magnetic approximation and the inline current feed solution
for the same geometry. We proceed as in the previous part, we choose $\phi_1=
H a_1 + c_{max}$. Remember that in the inline case, 
$\phi$ is a linear function by parts. $\forall x \in ]a_i,a_{i+1}[$, 
$$P_{i+1}(x)=(d_i\sin(P_i(a_i))+P_i^{\prime}(a_i))(x-a_i)+P_i(a_i).$$
Let us define:
$$\begin{array}{rcl}
\alpha_i &=& P_i^{\prime}(a_i)-f^{\prime}(a_i) = P_i^{\prime}(a_i)-H~, \\ 
\beta_i &=& P_i(a_i)-f(a_i)~.
\end{array}$$
We obtain that $\alpha_1=-\gamma/2$, $\beta_1=0$ and for a $n$ junction circuit
$\alpha_{n+1}= \gamma/2$. We estimate $\alpha_{i+1}$:
$$\alpha_{i+1}=d_i\sin(P_i(a_i))+P_i^{\prime}(a_i)-H=d_i\sin(P_i(a_i))+\alpha_i~,$$
we obtain, $\alpha_{i+1} = \sum_{j=1}^i d_j \sin(P_j(a_j)) + \alpha_1$, thus
\begin{equation}\label{alpha_magne}
|\alpha_{i+1}| \leq \sum_{k=1}^n d_k ~.
\end{equation}
We write $\beta_{i+1}$:
\begin{eqnarray*}
\beta_{i+1} &=& (d_i\sin(P_i(a_i))+P_i^{\prime}(a_i))(a_{i+1}-a_i)+P_i(a_i)-Ha_{i+1}
+b ~, \\
&=& (d_i\sin(P_i(a_i))+P_i^{\prime}(a_i)-H)(a_{i+1}-a_i)+P_i(a_i)-Ha_i+b ~, \\
&=& \alpha_{i+1}(a_{i+1}-a_i)+\beta_i ~.
\end{eqnarray*}
Thus if $\beta_1=0$ then 
\begin{equation}\label{beta_magne}
\beta_{i} = \sum_{k=1}^{i-1} \alpha_{k+1}(a_{k+1}-a_k)~.
\end{equation}
Now using the bounds on the $\alpha$'s and bounding the $l_i$'s we get
\begin{equation}\label{majo4}
|\beta_{i}| \leq n l_b \sum_{k=1}^n d_k~.
\end{equation}
This shows that the $\gamma_{max}$ of Eq.(\ref{e5.1}) tends to the magnetic 
approximation when $\sum_{k=1}^n d_k$ tends to $0$. 


\begin{thebibliography}{10}

\bibitem{josephson} 
{\sc B. D. Josephson}, Phys. Lett. {\bf 1}, 251, (1962).

\bibitem{Barone} {\sc A. Barone and G. Paterno}, {\em Physics and 
Applications of the Josephson effect}, J. Wiley, (1982).

\bibitem{Likharev} {\sc K. Likharev}, {\em Dynamics of Josephson junctions and
circuits}, Gordon and Breach, (1986).

\bibitem{Salez} {\sc M. Salez et al.}, Proc. SPIE Conf. on Telescopes
and Astronomical Instrumentation, Hawaii, 2002 (August 22-28),
col. 4855, p. 402, Proc. 4th European
Conference on applied superconductivity, EUCAS 99, 651, (1999)

\bibitem{cfv95} {\sc J. G. Caputo, N. Flytzanis and M. Vavalis}, 
{\em A semi-linear elliptic pde model for the static solution of Josephson 
junctions}, International Journal of Modern Physics C, vol. 6, No. 2, 
241-262, (1995).

\bibitem{cfgv96} {\sc J. G. Caputo, N. Flytzanis, Y. Gaididei and M. Vavalis},
{\em Two-dimensional effects in Josephson junctions: I static properties},
Phys. Rev. E, 54, No. 2, 2092-2101, (1996).


\bibitem{bcf02} {\sc A. Benabdallah, J. G. Caputo and N. Flytzanis}, 
{\em The window Josephson junction: a coupled linear nonlinear system},
Physica D, 161, 79-101, (2002).


\bibitem{cg04} {\sc J. G. Caputo and Y. Gaididei}, 
{\em Two point Josephson junctions in a superconducting stripline: static case.},
Physica C, 402, 160-173, (2004).

\bibitem{cl} {\sc J. G. Caputo and L. Loukitch}, {\em Dynamics of 
point Josephson junctions in microstrip line.}, Physica {\bf 425} (2005)
69-89.

\bibitem{mghg} {\sc J. H. Miller, Jr., G. H. Gunaratne, J. Huang, and
T. D. Golding}, Appl. Phys. Lett. 59, (25), 3330 (1991).

\bibitem{cftv03} {\sc J. G. Caputo, N. Flytzanis, A. Tersenov and
M. Vavalis}, {\em Analysis of a semi-linear pde for modeling static
solutions of Josephson junctions}, SIAM J. of Math. Analysis, 
{\bf 34}, 1356-1379, (2003).

\bibitem{uclocr95} {\sc A. V. Ustinov, M. Cirillo, B. H. Larsen, V. A.
Oboznov, P. Carelli, and G. Rotoli}, Phys. Rev. B 51, (5), 3081 (1995).


\bibitem{fgb92} {\sc R. Fehrenbacher, V. B. Geshkenbein and G. Blatter}, 
{\em Pinning phenomena and critical currents in disordered long Josephson
junctions}, Phys. Rev. B {\bf 45}, 5450, (1992). 

\bibitem{it9596} {\sc M. A. Itzler and M. Tinkham}, {\em Flux pinning in large
Josephson junctions with columnar defects}, 
Phys. Rev. B {\bf 51}, 435, (1995),
{\em Vortex pinning by disordered columnar defects in large 
Josephson junctions}, 
Phys. Rev. B {\bf 53}, 11949, (1996)

\bibitem{sbcld} {\sc M. Salez, F. Boussaha, J. G. Caputo and L. Loukitch}, 
{\em SQUID properties of non-uniform, parallel 
superconducting junction arrays}, In press. 

\bibitem{maple} http://www.maplesoft.com/

\end{thebibliography}
\end{document}